\documentclass[acmsmall]{acmart}
\AtBeginDocument{%
  \providecommand\BibTeX{{%
    \normalfont B\kern-0.5em{\scshape i\kern-0.25em b}\kern-0.8em\TeX}}}

\setcopyright{cc}
\setcctype{by}
\acmJournal{PACMHCI}
\acmYear{2025} \acmVolume{9} \acmNumber{7} \acmArticle{CSCW261} \acmMonth{11} \acmPrice{}\acmDOI{10.1145/3757442}

\usepackage{array}
\usepackage{enumitem}
\usepackage{multirow}
\usepackage{makecell}
\usepackage{soul,xcolor}
\setstcolor{red}
\usepackage{longtable}

\usepackage{subcaption}
\usepackage{colortbl}
\usepackage{caption}

\usepackage{xspace}
\usepackage{longtable}

\begin{document}

\title{``\systemname, How to Let The Lettuce Dry Without A Spinner?'': Exploring User Perceptions of Using An LLM-Based Conversational Assistant Toward Cooking Partner}
\renewcommand{\shorttitle}{Exploring User Perceptions of Using An LLM-Based Conversational Assistant Toward Cooking Partner}



\author{Szeyi Chan}
\authornote{Both authors contributed equally to this research.}
\email{chan.szey@northeastern.edu}
\affiliation{%
  \institution{Northeastern University}
  \country{USA}
}

\author{Jiachen Li}
\authornotemark[1]
\email{li.jiachen4@northeastern.edu}
\affiliation{%
  \institution{Northeastern University}
  \country{USA}
}

\author{Bingsheng Yao}
\email{arthuryao33@gmail.com}
\affiliation{%
 \institution{Northeastern University}
 \country{USA}
 }

\author{Amama Mahmood}
\email{amahmo11@jhu.edu}
\affiliation{%
 \institution{Johns Hopkins University}
 \country{USA}
 }

\author{Chien-Ming Huang}
\email{chienming.huang@jhu.edu}
\affiliation{%
 \institution{Johns Hopkins University}
 \country{USA}
 }

\author{Holly Jimison}
\email{h.jimison@northeastern.edu}
\affiliation{%
  \institution{Northeastern University}
  \country{USA}
  }

\author{Elizabeth D Mynatt}
\email{e.mynatt@northeastern.edu}
\affiliation{%
  \institution{Northeastern University}
  \country{USA}
}

\author{Dakuo Wang}
\authornote{Corresponding author d.wang@northeastern.edu}
\email{d.wang@northeastern.edu}
\affiliation{%
  \institution{Northeastern University}
  \country{USA}
}

\renewcommand{\shortauthors}{Szeyi Chan et al.}


\newcommand{\systemname}{\textit{Mango Mango}\xspace}
\newcommand*{\revised}{\textcolor{red}}
\newcommand*{\revisedfinal}{\textcolor{black}}
\begin{abstract}
The rapid advancement of Large Language Models (LLMs) has created numerous potentials for integration with conversational assistants (CAs) assisting people in their daily tasks, particularly due to their extensive flexibility. However, users' real-world experiences interacting with these assistants remain unexplored. 
In this research, we chose cooking, a complex daily task, as a scenario to explore people's successful and unsatisfactory experiences while receiving assistance from an LLM-based CA, \systemname. 
We discovered that participants value the system's ability to offer customized instructions based on context, provide extensive information beyond the recipe, and assist them in dynamic task planning. 
However, users expect the system to be more adaptive to oral conversation and provide more suggestive responses to keep them actively involved. Recognizing that users began treating our LLM-CA as a personal assistant or even a partner rather than just a recipe-reading tool, we propose five design considerations for future development.
\end{abstract}

\ccsdesc[500]{Human-centered computing~User studies}
\ccsdesc[500]{Human-centered computing~Sound-based input / output}
\ccsdesc[500]{Human-centered computing~Auditory feedback}
\ccsdesc[500]{Human-centered computing~Empirical studies in HCI}

\keywords{user study, exploratory study, large language model-based conversational assistant}



\maketitle

\section{Introduction}
Current conversational assistants (CAs), such as Amazon's Alexa, Apple's Siri, and Google Assistant, are important in our daily lives, especially in home-based settings ~\cite{10.1145/3196709.3196772, hoy2018alexa,allouch2021conversational}. The ``hands-free'' and ``eyes-free'' design enables users to effortlessly access information through voice commands for simple question-answering tasks, including setting reminders, providing weather updates, and searching for recipes ~\cite{mathur2022collaborative, 10.1145/3313831.3376225, zubatiy2021empowering}.

Nevertheless, CAs face limitations and challenges when instructing users with hands-on tasks in family-centered scenarios, such as experimenting with new recipes for special family dinners, resolving urgent plumbing problems, or collaboratively assembling new furniture. These tasks often require fundamental knowledge in areas in which family members may not always have expertise, leading them to seek guidance through online instructional videos or product manuals~\cite{kraus2020successful, bercher2021yourself}. In particular, cooking requires steps like preparing food, finding ingredients, measuring the correct amount, and planning, all while the cook's hands are occupied with food preparation~\cite{neumann2021recipe,craik2006planning, nouri2019supporting}. Unfortunately, current CAs cannot provide comprehensive and continuous support with these tasks~\cite{stolwijk2022increasing}. Existing CAs rely on predefined dialogue logic and often struggle with language comprehension, prohibiting natural back-and-forth conversations for complex tasks~\cite{ammari2019music, arnold2022does, stolwijk2022increasing, cho2019once, 10.1145/3098279.3098539, 10.1145/2858036.2858288, 10.1145/3313831.3376760}. Therefore, exploring new approaches is essential to effectively address these challenges and enhance the support provided by CAs in such complex, interactive scenarios.

Recent advancements in language models, particularly large language models (LLMs), for example, GPT-3.5/4~\cite{OpenAI2023GPT4TR}, LLaMA~\cite{touvron2023llama}, and PaLM~\cite{chowdhery2022palm}, show the ability to overcome the limitations of language models used in current CAs.  Existing works have shown LLMs have natural language understanding (NLU) ~\cite{allen1995natural} and generation (NLG) ~\cite{reiter1997building, semaan2012natural} capabilities to understand users’ lengthy text input and accommodate multi-turn dialogues~\cite{xu2023leveraging}. Despite significant advancements, the integration of LLMs into CAs for real-world scenarios remains underexplored. Specifically, there is limited understanding of whether CAs with LLMs can address key challenges faced by traditional CAs, such as adapting to diverse user needs and supporting complex, domain-specific tasks like cooking or collaborative problem-solving. While LLMs are good at understanding and generating text, their practical application in CAs requires further explorations to tailor responses to dynamic contexts and ensure usability across diverse interactions.

To explore the integration of LLMs into CAs, our research consists of two parts: 1) developing an LLM-based system, \systemname, specifically tailored to help individuals cook at home and 2) conducting a mixed-method in lab exploratory study to evaluate users' experiences through preparing for a salad.
Following the study, we performed both qualitative and quantitative research analyses with semi-structured interviews, surveys, and system logs. 
Our research is guided by two primary questions: \textbf{(1) How do users perceive LLM-based CA in cooking scenarios through their interaction experiences?} and \textbf{(2) What are the design implications of LLM-based CAs aimed at assisting users in real-world practices like cooking? }

The questionnaire results from the study indicate that participants generally have a positive experience using \systemname. Users' feedback from the interviews shows appreciation for features including receiving aid beyond the recipe, recollection of the current cooking status, personalized instructions, task planning, free control of the cooking process by user preference, etc. However, some design aspects require improvement, including managing information overload from responses, addressing issues with understanding oral expressions, minimizing redundant interactions with the system, facilitating more engaging dialogues with the CAs, and more. Additionally, the study found that users' perceptions of \systemname changed during their interaction, from perceiving CAs simply as a tool, to a personal assistant, and to a partner. Based on the findings, we discussed design considerations for leveraging LLMs' NLU and NLG capabilities to enhance the effectiveness and usability of LLM-based CAs specifically in cooking applications.

The main contributions of our paper are summarized as follows:\hfill
\begin{enumerate}[noitemsep,topsep=0pt]
\item We developed a conversational assistant system that integrates LLM (GPT 3.5-Turbo) to guide users in cooking scenarios. 
\item We conducted a mixed-methods exploratory study with 12 participants in a home kitchen setting to better understand user experiences when using LLM-based CAs in cooking tasks.
\item We summarized the key themes of successful and unsatisfactory user experiences based on semi-structured interviews.
\item We provided design implications for future LLM-based CAs in cooking scenarios.

\end{enumerate}

\section{Related Work}
We first focus on recent developments in CAs designed to meet real-world demands in Section~\ref{sec:relate-ai4human}. Additionally, we discuss the evolution of language models and recent applications developed with LLM in Section~\ref{sec:relate-llm}. 
Lastly, we touch on existing work that utilizes AI techniques to enhance cooking scenarios in Section~\ref{sec:relate-ai4cooking}.

\subsection{CAs for Human: Real-World Challenges}
\label{sec:relate-ai4human}
Researchers have been exploring using CAs with language models in real-world situations to assist people in accomplishing daily tasks. CAs applications like chatbots~\cite{10.1145/3290605.3300439, 10.1145/3381804, 10.1145/3544548.3581252, han2022faq, ayers2023comparing, 10.1145/3581641.3584031} have been developed and tested to successfully assist people in completing various activities. For example, smart CAs have shown promising capability as reliable healthcare technologies for elders~\cite{pradhan_use_2020, brewer_empirical_2022, berube_reliability_2021, harrington_its_2022, bartle_second_2022, bartle_this_2023}. CAs are also used for other scenarios, such as travel~\cite{poran2022with, cambre_firefox_2021}, music~\cite{ammari2019music}, education~\cite{10.1145/3491102.3517479, 10.1145/3585088.3589354, chubb2022interactive, 10.1145/3411764.3445039, 10.1145/3078072.3084330, jakesch2023co, 10.1145/3459990.3465195, 10.1145/3381002}, home~\cite{beirl2019using, 10.1145/3313831.3376344, 10.1145/3264901, 10.1145/3196709.3196772}, etc., showing promising utilities~\cite{10.1145/3411764.3445536, 10.1145/3173574.3173869, 10.1145/3313831.3376225}.

However, challenges in developing CAs are identified primarily due to disparities in user perceptions of the system's capabilities. Issues like speech detection failure and faulty recognition can occur~\cite{pearl2016designing, myers2018patterns}. The use of heuristics in most existing commercial CAs limits the scope of questions that can be answered and constrains the support of basic interaction functionalities (e.g. setting reminders), which can potentially cause users to feel discouraged and lower their expectations of the technology’s capabilities~\cite{ammari2019music, cho2019once, arnold2022does, 10.1145/3098279.3098539, 10.1145/2858036.2858288, 10.1145/3313831.3376760}. Additionally, current CAs face challenges in responding to queries about external sources, lapses in providing comprehensive details, and lack of ability to provide broader context ~\cite{hwang2023rewriting, le2023improved}.~\citet{jaber2024cooking} highlights the challenges and importance of context awareness when working on complex tasks such as cooking. Current commercial VAs often fail due to a lack of contextual awareness, leading to irrelevant responses. This underscores the need for developing CAs that can maintain and use shared context during the interaction to improve interaction quality and task support.

The aforementioned limitations are related to LM-based CA, and the advancement of LLM offers the potential to effectively address and mitigate these issues. To unlock the potential of LLM, previous work explored that designing effective prompting~\cite{xiao2023supporting} and facilitating information retrieval within conversational contexts~\cite{liao2020conversational} would provide natural user experiences. However, the question of how people adapt these benefits of LLM with CAs in real-life tasks remains an important yet unexplored topic. Our research aims to fill the gap in exploring user experiences using LLM-based CAs, focusing on cooking in a home kitchen setting, which we will describe the rationale for and previous work on in the next section.

\subsection{Leveraging the Potential of LLM in Everyday Applications}
\label{sec:relate-llm}
Current language models require substantial amounts of data for training, facing challenges like fine-tuning a system to generate responses with varying tones ~\cite{alfifihowdy}. However, innovative methods and algorithms, such as instructional-finetune~\cite{wei2021finetuned, chung2022scaling} and reinforcement learning with human feedback (RLHF) ~\cite{christiano2017deep, ouyang2022training}, have revolutionized the potential of LLMs such as LLaMA~\cite{touvron2023llama}, FLAN~\cite{wei2021finetuned, chung2022scaling}, PALM~\cite{chowdhery2022palm}, InstructGPT~\cite{ouyang2022training}, and GPT-4~\cite{ OpenAI2023GPT4TR}. These models are fine-tuned on various natural language tasks, enabling them to effortlessly comprehend all instructions and generate high-quality text content~\cite{brants2007large, omar2023chatgpt, robinson2023leveraging}. LLMs also show the ability to handle lengthy text input (e.g., GPT-4~\cite{OpenAI2023GPT4TR} can take $32,000$ tokens) to perform tasks that traditional LMs cannot handle, such as multi-turn conversations.

As LLM technology advances, researchers are actively exploring various potential applications~\cite{wang2023enabling, wu2022ai, liu2022will, lee2022coauthor, kim2022stylette, jiang2022promptmaker, li2023chatdoctor, swanson2021story, lee2023dapie, hasan2023sapien}. Researchers are particularly interested in utilizing LLM's ability to process inputs through prompt engineering and generate outputs that combine extensive dataset knowledge to make these applications come true~\cite{dang2022prompt}, including qualitative analysis with cultural context comprehension~\cite{xiao2023supporting}, connecting LLM to robots for executing complex real-world tasks with task planning~\cite{ahn2022i}, co-creating tools for story and sketch generation~\cite{chung2022talebrush}, software engineering tools for code generation~\cite{jiang2022discovering}, and tools for mental health awareness~\cite{kumar2023exploring, xu2023mentalllm}.

However, an underexplored area remains in utilizing LLMs for everyday home-based tasks, such as cooking. Our work aims to leverage the advantages of LLM technology and incorporate conversational assistance to bridge the gap between LLM capabilities and the lack of consideration for system design implications from a human-computer interaction perspective in everyday scenarios.

\subsection{AI for Cooking}
\label{sec:relate-ai4cooking}
Cooking is a common daily task that requires the execution of sequential steps and multitasking skills to enhance efficiency~\cite{neumann2021recipe,craik2006planning, nouri2019supporting}. Individuals new to cooking or attempting to prepare a new recipe often turn to resources like cookbooks and YouTube videos for guidance~\cite{logie2010multitasking, kosch2019digital,weber2023designing}. Their hands are usually occupied during the cooking process, restricting their capability to gather and process information. Various AI cooking assistants have emerged to address the challenge by using multiple modes of communication, including text, video, and audio, across various devices such as screens, tablets, and computers~\cite{sato2014mimicook, chen2010smart}. For instance, AI-powered cooking assistants like ``Cooking Nav’’ ~\cite{hamada2005cooking}, ``AskChef’’~\cite{nouri2019supporting}, and ``MimiCook’’~\cite{sato2014mimicook} provide multi-tasking planning, step-by-step guidance, and interactive ingredient weight projections. These tools help individuals optimize their cooking process and maintain their hands during the cooking process.

While there has been an increase in the number of AI-powered cooking assistants available, many remain limited to providing guidance strictly based on pre-set recipes and relying on multimodal inputs for context~\cite{sato2014mimicook, chen2010smart,hamada2005cooking,neumann2021recipe}. Researchers explore using smart CAs for cooking assistants, but traditional language models and pre-determined heuristics may limit their flexibility, ability to answer questions, and multi-turn conversation capacity~\cite{winkler2019alexa, zhao2022rewind}. Our study explores the potential of LLM-powered cooking CAs for a seamless, interactive cooking experience through voice commands, user experience, and design considerations for further development, allowing users to complete tasks at their own pace and receive immediate assistance.

\section{Methods}
To attain insights into users' expectations and feedback during interactions with LLM-CA and to frame design suggestions that best utilize the unique strengths of LLMs in real-world scenarios, we conducted an exploratory user study utilizing a mixed-methods approach. This section presents an overview of our approach, including the implementation details of the system we developed and the user study specifics.

We will first explain the development and design of our LLM-based CA system, providing a detailed overview of the system pipeline and prompt design in Section~\ref{sec:method-system_design}. This will be followed by a discussion of the experiment design and procedure in Section~\ref{sec:method-study_design}. Next, we will describe the recruitment process and participant demographic information in Section~\ref{sec:recruitment_process} and~\ref{sec:demographic}. We will then outline the data collection methods in Section~\ref{sec:method-data_collection}, including semi-structured interviews, surveys, and system logs. Finally, we will detail the analysis process in~\ref{sec:data_analysis}.

\subsection{System Design}
\label{sec:method-system_design}
\begin{figure}[t]
  \centering
  \includegraphics[width=.98\linewidth]{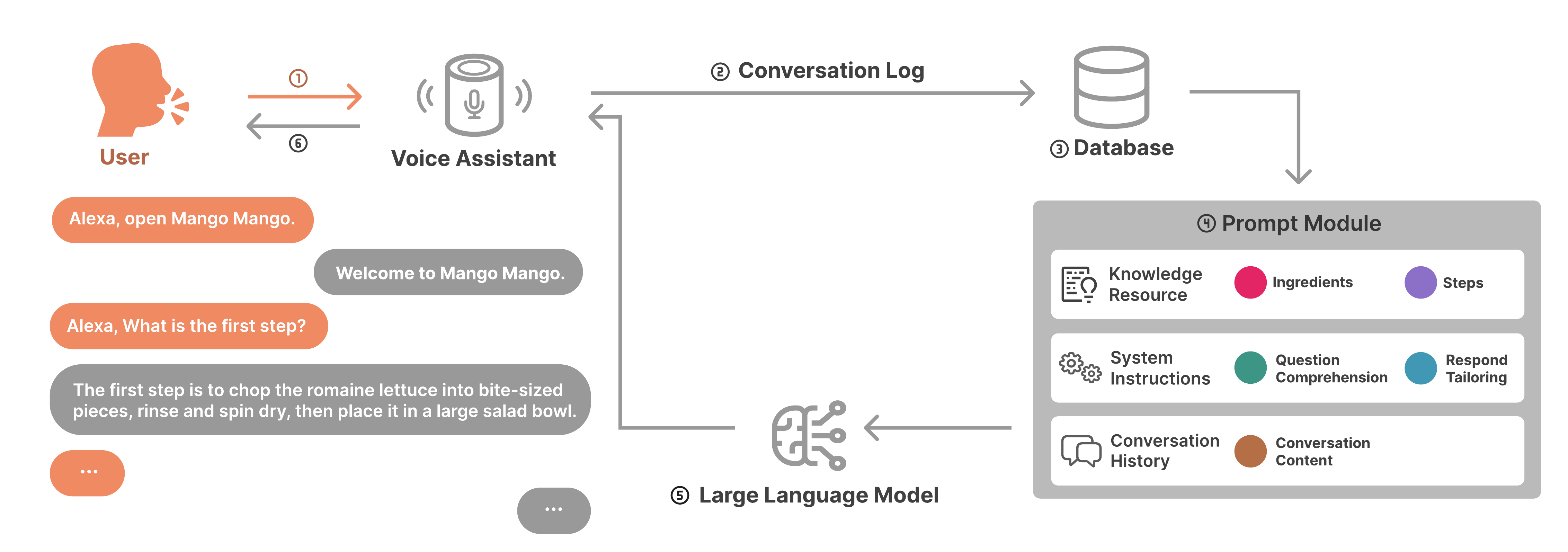}
  \caption{Simplified system diagram of \systemname. The flow of the system is as follows: (1) Users speak to Alexa as voice input, then the text-to-speech process; (2) The transcribed inputs are saved in the conversational log (database); (3) The conversational log, along with the updated conversational history, was processed in the prompt module. The prompt module included knowledge resources, instructions, and conversation history; (4) The completed prompt is then sent to the GPT-3.5 Turbo; (5) The resulting response is sent back to Alexa; (6) Finally, Alexa converts the response into speech to the user to complete the system loop.}
  \label{fig:system_diagram}
  \Description{This is the simplified system diagram of \systemname. The flow for the system diagram starts from the left to right and then circles back to the left. First, the user provided voice input to Alexa. The flow then will start to move to the right. Then, an arrow to the top for the second step, where the text conversion process is saved in the conversational log. It has a data storage icon here. Then the flow goes around to step 3. At this step, the conversational log and the updated conversational history were processed in the prompt module. The prompt module included knowledge resources, instructions, and conversation history. The flow is now in the 6'o 6-clock position. At this step, the generated message is sent to the GPT-3.5 Turbo for processing. Finally, the flow is back to the left. The resulting response is sent back to Alexa and converted into speech to complete the system loop}
  \vspace{-0.15in}
\end{figure}

\begin{figure}[t]
  \centering
  \includegraphics[width=\linewidth]{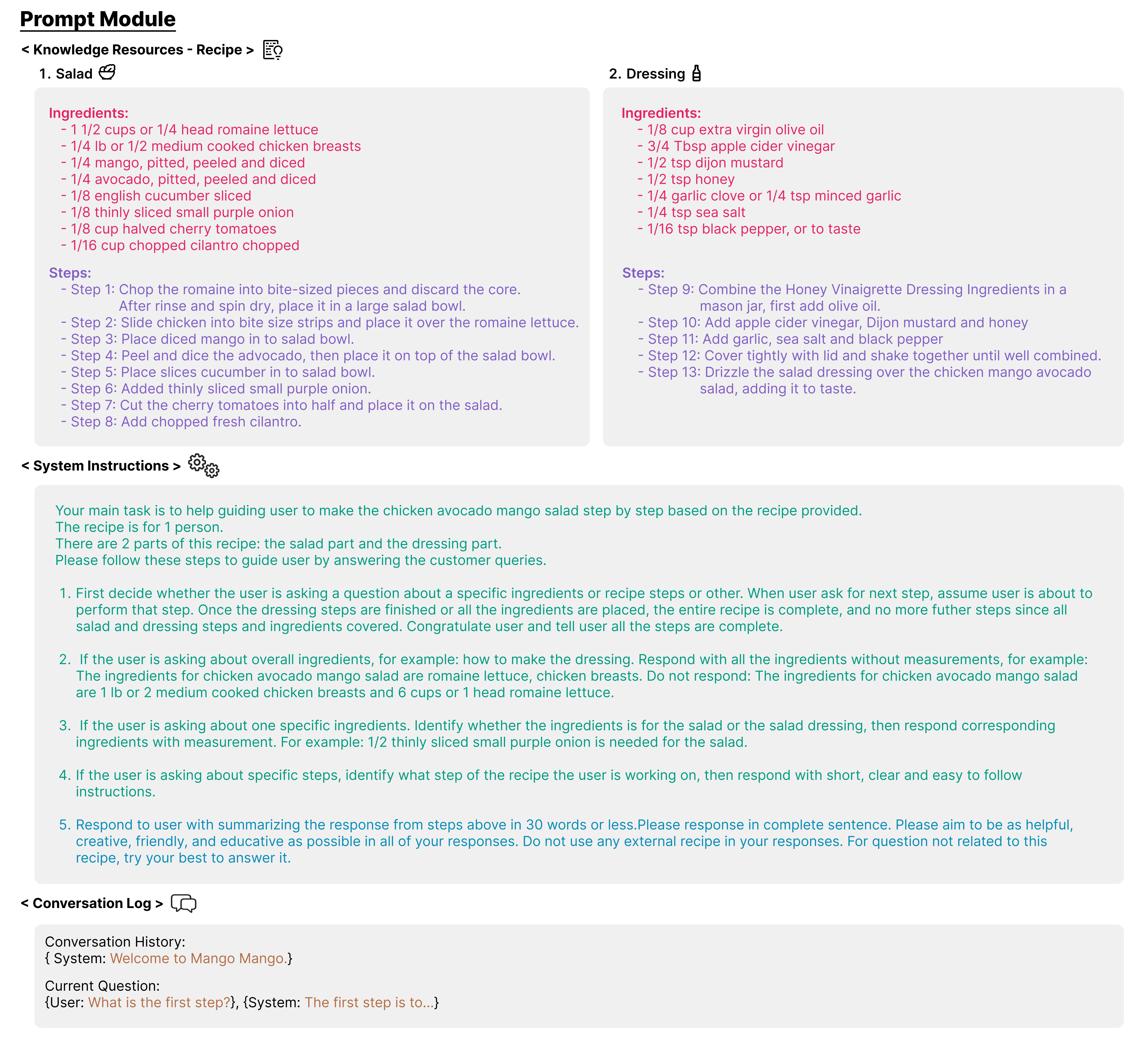}
  \caption{Detail components of \systemname’s prompting module. The prompting module contains the instructions module (left) and the knowledge resources (right). The instructions module will understand the users’ input from the conversation, then the model selects the appropriate knowledge resource based on the user's input. The Knowledge Resources cover all the necessary information related to the recipe, including ingredients and steps. Finally, return the tailored guidance or suggestions based on users’ inquiries.}
  \Description{This is the detailed Components of \systemname’s prompting module. It divided into two main categories: Instructions and Knowledge Resources.Under Instructions on the left:There are three parts indicating the flow of information intake. First one is the User Input, including History Conversation and Current Question. Secondly is the Question Comprehension. Thirdly is Response Tailoring, with reorganizes and personalizes its responses, aiming to create human-like narratives. Under Knowledge Resources on the right: There are two sets of information. The first set contains General Ingredients with special Ingredients (with a note mentioning ``dressing, etc.'').The second set is labeled Steps included special steps. Finally with a dotted line connects Instructions and Knowledge Resources.}
  \label{fig:prompt_design}
  \vspace{-0.15 in}
\end{figure}

In this study, we introduced an LLM-based CA, \systemname, designed to assist users in completing a recipe. 
We selected Amazon Alexa as our voice-based CA and used the Alexa skill platform because of its flexible functionality and built-in features, particularly the text-to-speech conversion technology~\cite{lopatovska2019talk,liew2023alexa}. Moreover, we have integrated it with the GPT-3.5-Turbo model, which has elevated its natural language processing capabilities. Figure~\ref{fig:system_diagram} demonstrates the complete pipeline of our system.

\subsubsection{Alexa Skill}

The Alexa Skill Kit is a development framework for CA applications that can be integrated into Amazon smart speakers, such as Amazon Echo and Dot. 
This framework leverages Amazon's fundamental natural language and speech recognition technologies, such as Text-to-Speech (TTS),  Speech-to-Text (STT), and intent recognition, to enable necessary speech recognition and text conversion functionalities for CAs, and allow users to customize the back-end application pipelines with a significant degree of freedom.

When a user activates the skill using a predefined invocation name, Amazon's STT technology converts the user's spoken queries into text. The text is then sent to the backend of the skills, where it undergoes processing through our LLM system, as discussed in Section~\ref{sec:method-system_design-llm}. Once the LLM has generated a response, it is sent back through the API and converted to synthetic voices using Amazon's TTS technology. The system awaits further user inputs after providing the response. 


\subsubsection{LLM Selection}
\label{sec:method-system_design-llm}

Our LLM-CA system utilizes OpenAI's GPT-3.5-Turbo LLM in the backend. The selection of an LLM was guided by an evaluation of several key factors. Firstly, GPT-3.5-Turbo has demonstrated proficiency in both NLU and NLG, making it the backbone of the web-based chat assistant, ChatGPT. Furthermore, its capability to manage extensive input content enables us to send numerous previous rounds of conversation histories simultaneously, resulting in more coherent and suitable multi-round conversations.

Secondly, GPT-3.5-Turbo provides comprehensive and stable API support, which is crucial to supporting the smoothness of our lab experiments. 
During the implementation of our system, we endeavored to utilize the GPT-4, a more advanced LLM, which boasts superior capabilities to its predecessor, GPT-3.5-Turbo. Regrettably, despite its acclaimed superiority, we observed a suboptimal response time from GPT-4 API, making it more prone to exceeding the Alexa Skill's backend waiting time limit. This caused the Alexa Skill to be forcibly terminated before the response from GPT-4 was generated and sent back.
In summary, GPT-3.5-Turbo is an ideal LLM benchmark to provide stable support while providing the unique advantages of LLMs over traditional language models for our exploratory study.

After the user’s voice input is captured and correctly recognized by the Alexa Skill, it is converted into text and sent to a database for conversation log storage. The conversation log is then forwarded to the back-end prompting module, where the input text is organized and reconstructed into a complete query. This query is sent to the LLM, GPT-3.5-Turbo in our implementation, via API to generate a response.

\subsubsection{Database}
To effectively maintain and manage the conversation log, we incorporated a database into our system design, a method similarly utilized in previous studies~\cite{mahmood2024user,yang2024talk2care}. In our implementation, we integrated a shared Google Sheet as a middleware database. Each interaction is logged in real-time and stored in the Google Sheet, ensuring accessibility, transparency, and efficient tracking of conversational data. This logged data is dynamically forwarded to the prompt module, where it becomes part of the prompt input. By including conversation history in the prompt, the system leverages prior conversations to enhance contextual understanding to generate more relevant and personalized responses.

\subsubsection{Prompting Module}
Our prompting module is specifically designed to support the cooking scenario with recipes. In addition to the conversation log, it is structured around two additional core components: \textbf{Knowledge Resources} and \textbf{Instructions}. Figure~\ref{fig:prompt_design} illustrates the complete prompt, developed based on the salad recipe used in our lab experiment. To ensure the system’s functionality, we conducted three pilot studies within the research team to iteratively test and refine the system. These studies helped identify key areas for improvement and informed the final design of the module.

\paragraph{\textbf{Knowledge Resources}} \label{Knowledge Resources}

In the cooking scenario used in our experiment, the Knowledge Resources were structured based on a design choice to ensure clarity and scalability. These resources consisted of two primary components: ingredients and steps, encompassing all necessary information related to the recipe. During our pilot study, we observed that GPT struggled to distinguish whether an ingredient and steps were intended for the salad or the dressing. Therefore, subcategories were created within each component to clarify distinct elements of more complex recipes. For example, a subsection for ``salad'' and ``dressing'' were added under the categories in our experiment, as its preparation was relatively independent of the main salad preparation process. This structure reflects a design decision aimed at maintaining flexibility and usability. By organizing the recipe in additional subcategories, the framework supports effortless scaling to accommodate various recipes, regardless of complexity, while ensuring that the information remains logically structured for users.

Regarding the data resources used to create the cooking steps, existing work~\cite{weber2023designing} has discovered that people tend to search for recipes on the internet in real-life scenarios, especially YouTube recipe teaching videos. Therefore, we chose YouTube cooking teaching videos as recipe source data in our system. Specifically, we first transcribed the video to capture the ingredient list and each cooking step, ensuring the instructions' originality and identicality to the instructions from the video. We leveraged bullet points to list each individual ingredient information, such as the name and quantity of the ingredients required for the recipe, as well as individual step details, so that the LLM can more conveniently and accurately locate the sequence of instructions and the details of each item. Additionally, the distinct component in this recipe was the dressing of salad, so we separated the list of ingredients for the dressing into a subsection of special ingredients. The process of transferring YouTube videos into the Knowledge Resources that could be used in our system is highly replicable and scalable to different recipes. For our experiment, we selected the chicken avocado mango salad, and we will delve into the recipe specifics in Section~\ref{sec:method-study_design}.
  
Cooking tasks vary widely due to the unique requirements of different recipes. The structures of Knowledge Resources were designed to be manually adaptable, catering to various recipes like sandwiches, cocktails, or no-bake desserts, in line with the procedures outlined in the earlier section. While the content of ingredients and cooking steps differ, the method of inputting relevant information into these sections in the Knowledge Resources remains consistent. However, we still recognize that some unique recipes and scenarios might pose challenges in scaling up \systemname, which will be discussed in the limitations section.

\paragraph{\textbf{Instructions}}

LLMs possess an exceptional natural language generation ability and access an almost boundless wealth of knowledge, enabling them to answer a wide range of questions. However, LLMs also pose the difficulty of limiting the content they produce, which can lead to information overload, as highlighted in previous studies~\cite{hwang2023rewriting,corbett2016can,kim2021designers}. To optimize the LLM's natural language capabilities for cooking-related inquiries, we designed a detailed instruction pipeline in the prompting module, allowing for proper information retrieval from the Knowledge Resource to generate accurate responses based on requests. This comprises question comprehension and two aspects of response customization, namely recognition and targeted adaptation for different question types, as well as guidance on generating content more akin to human conversation.

We understand that cooking questions from users need different levels of detail and response methods. When users ask about necessary ingredients, it can be challenging to provide every detail, such as names, quantities, and specifics. Instead of overwhelming them with too much information, providing a list of ingredients is more effective.
If users want details about a particular ingredient or step, they should ask additional questions. The model should provide specific responses based on the knowledge available in the knowledge resources module.
The model should do more than just provide recipes. It should also respond to non-recipe-related inquiries. For example, users might ask practical questions about the cooking process, such as how to use kitchen tools or convert measurement units.

Therefore, as shown in the left module in Figure~\ref{fig:prompt_design}, we first require the model to understand user's input. Based on different users' inquiries, we provide targeted response guidance and suggestions for the model. 
When the user inquires about the required ingredients, we instruct the model to provide only a list of ingredients without specifying the quantity. We also provide a response template for such queries. 
If the user wants to know specific details about an ingredient, such as quantity, weight, or measurement conversion, the model needs to identify whether the user is referring to dishes or seasonings. Based on the user's input, the model selects the appropriate knowledge resource to provide an accurate response. This design can correctly identify and respond to user inquiries about specific ingredients in dishes that share common ingredients with condiments.
In regard to recipes, it is imperative that users receive descriptive and succinct guidance when inquiring about a specific step.

In addition to our tailored guidance for different question types, we've compiled some general tips to enhance the naturalness of the AI-generated responses. Our analysis revealed that the model often produces verbose and redundant content, which can overwhelm the recipient and disrupt the exchange's coherence. Furthermore, it is important to note that in certain instances, despite the fact that the LLM's response is comprehensive, Alexa Skill may truncate extended responses when speaking back to the user, leaving them incomplete mid-sentence. It is imperative to ensure that responses remain concise in order to avoid this issue.

As a result, we asked the model to prioritize brevity, aiming for responses that are no longer than 30 words, whenever feasible.
We require the model to limit its scope to answering only recipe-related questions from the given knowledge resource. However, when the user's questions exceed the boundaries of the recipe itself, we expect the model to leverage its world knowledge to provide comprehensive guidance.

\subsection{Experiment Design and Procedure}
\label{sec:method-study_design}
\begin{figure}[t]
    \centering
    \setlength{\belowcaptionskip}{-6pt}
    \subfloat[]
  {
    \label{fig:smarthome}
    \includegraphics[width=0.48\textwidth]{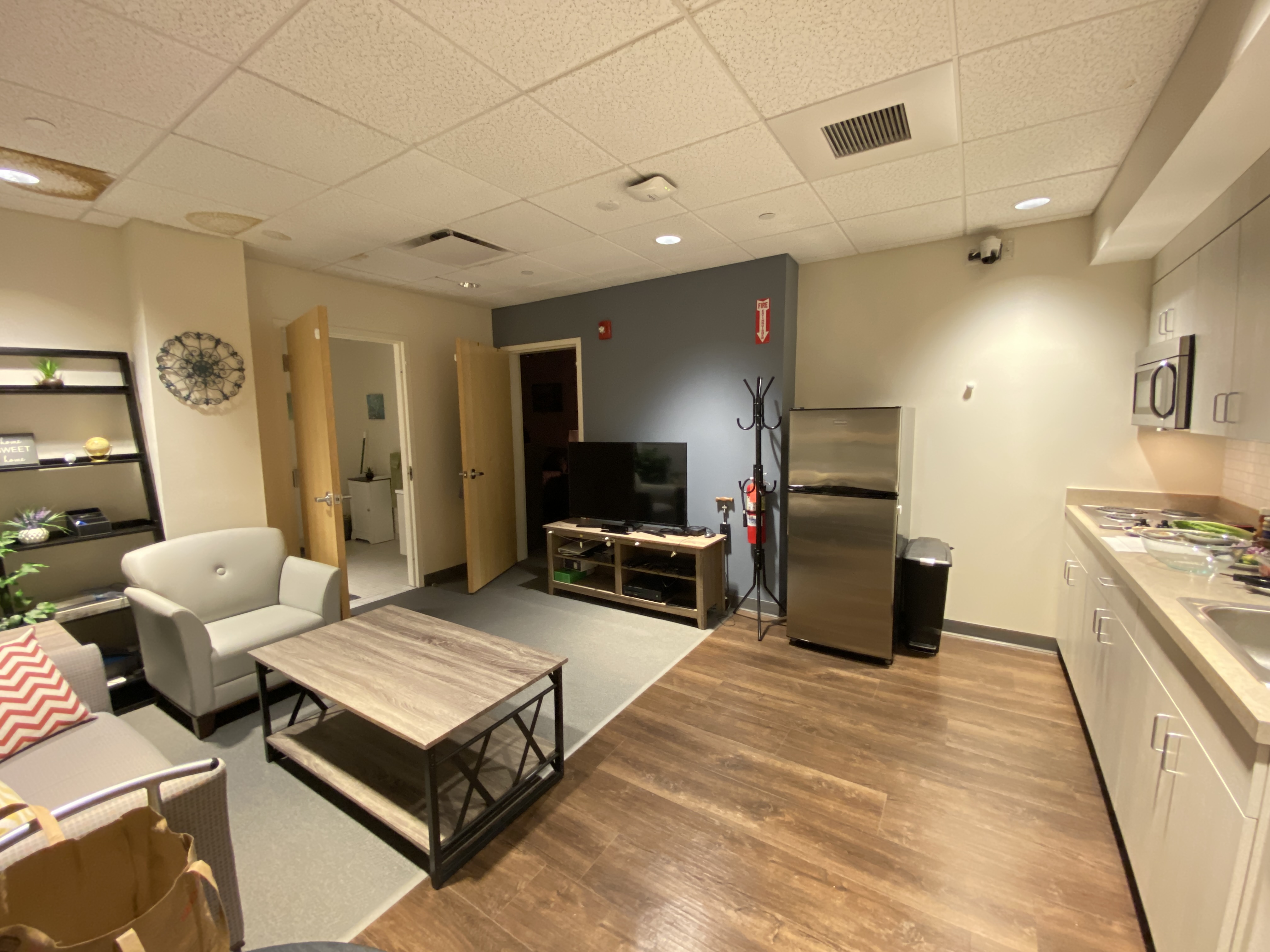}
  }
    \subfloat[]
    {
    \label{fig:participant}
    \includegraphics[width=0.48\textwidth]{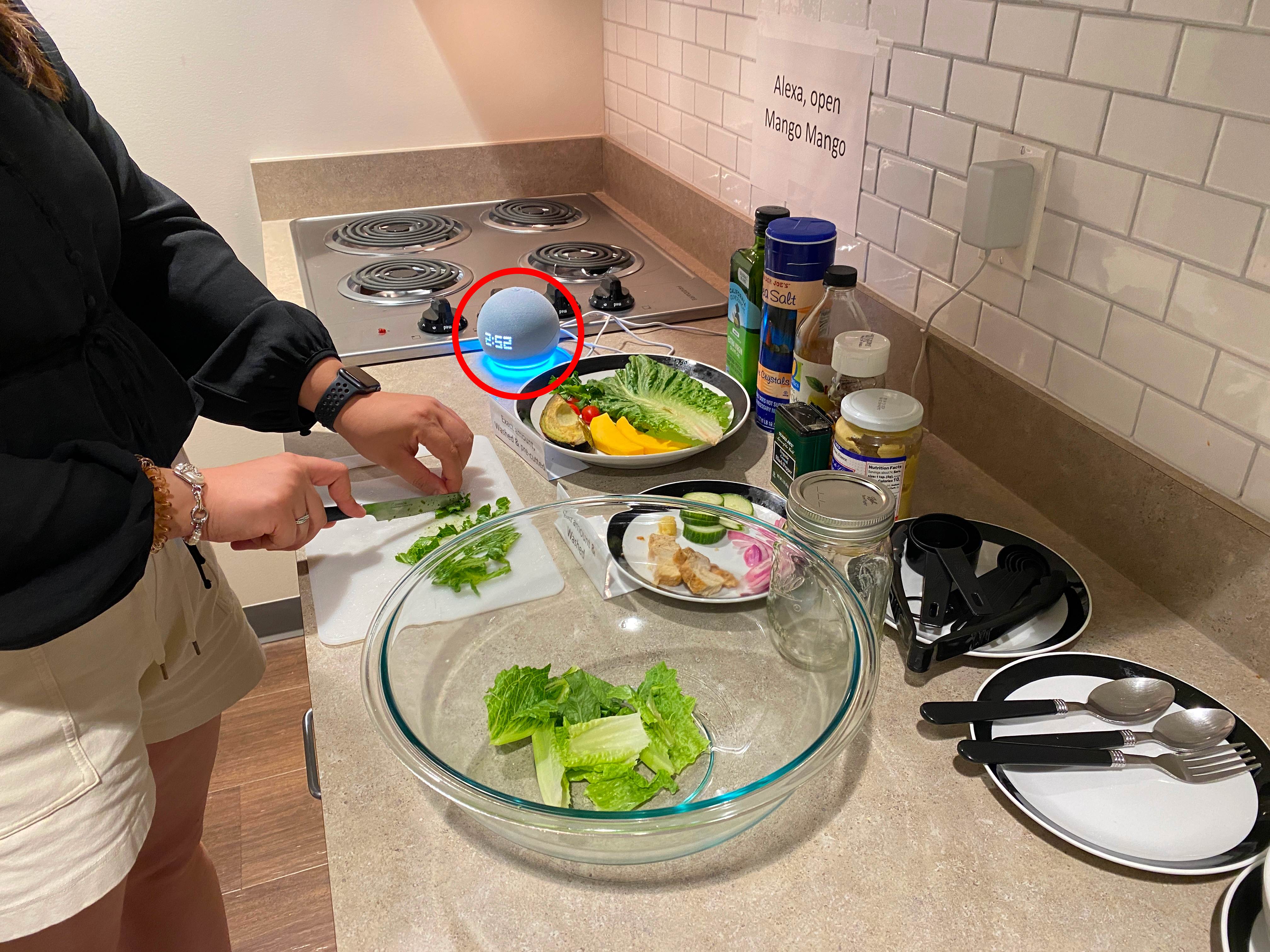}
  }

  \caption{Our study took place at the smart home laboratory (a). It was designed with a one-bedroom apartment floor plan with a fully functional kitchen and monitoring cameras. (b) Picture of a participant working on the experiment in the kitchen. Alexa is marked with a red circle on the left side of the table.}
  \label{fig:setup}
  \Description{Picture A is a picture shows the smart home laboratory designed with a one-bedroom apartment floor plan that includes a fully functional kitchen equipped with monitoring cameras. The angle is from the left of the room, the right side is the open kitchen, and a refrigerator is all on the other side of the room. There is a couch on the left side of the picture. In front of the couch is a small desk and a one-person sofa next to it. There is a shelf with decorations. Next to it is a bathroom with the door open and the bedroom with the door open. is a picture of the experiment area in the kitchen. Picture B is a picture of a participant during the experiment with the participant standing there cutting the lettuce. The picture shows the kitchen table arrangement for the experiment, with all the ingredients in place:  olive oil, apple cider vinegar, Dijon mustard, honey, salt, pepper, avocado, mango, lettuce, cherry tomato, cucumber, chicken, onion, some forks and knives, a salad bowl, and measurement spoons. Alexa is marked with a red circle on the left side of the table.}
  \vspace{-0.1in}
\end{figure}

To gain insights into users' experiences while interacting with \systemname during cooking, we organized an in-lab user study to simulate real-world scenarios and collect valuable feedback. Our study took place in a smart home laboratory, designed with a one-bedroom apartment floor plan that included a fully functional kitchen and equipped with monitoring cameras, as illustrated in Figure~\ref{fig:smarthome}. Participants should have completed demographic questionnaires and relevant surveys as part of the initial screening process. Upon arrival, researchers provide participants with an informative sheet detailing the data collection method and data storage and the participant protocol for the study. The researcher then asked for verbal consent from participants regarding the recording of their participation during the experiment. Subsequently, participants received a tutorial session guided by the research team. This session included a brief tour of the kitchen space and a trial interaction with \systemname to familiarize them with the Alexa voice assistant. Following this, participants viewed instructional YouTube videos that demonstrated how to prepare a chicken mango avocado salad. They were not required to memorize the video content but were encouraged to become familiar with the recipe. After viewing the video once, participants no longer had access to it, relying solely on our system, \systemname, for assistance when needed. Figure~\ref{fig:setup} shows the tabletop setup for the experiment. They then proceeded to prepare the salad while freely interacting with \systemname, without intervention from the researchers. Throughout this process, researchers observed the interactions from the control room and collected video recordings. Upon completing the dish, participants engaged in semi-structured interviews and surveys to reflect on their experiences. Our experiment was approved by the university's Institutional Review Board. To minimize any potential risks during the cooking process, we intentionally excluded using ovens, sharp knives, stoves, or any other appliances and tools that could threaten the participants' safety.

\subsubsection{Rationale}
We opted to utilize YouTube videos as our primary data resource for the following reasons. YouTube videos are immensely popular due to their detailed descriptions and rich visual cues. However, they lack voice interaction and sometimes require manual touch and scrolling for video control. On the other hand, voice assistants support hands-free interaction but may lack detailed information. Recognizing this disparity, we divided the use of these two tools into two phases: watching a video \textbf{before} cooking and interacting with the cooking assistant \textbf{during} the cooking process. Consequently, as previously described, we developed a workflow to translate video content into prompts. The full prompts can be found in Tables \ref{tab:prompt1} and \ref{tab:prompt2} in the appendix. In the user study, following this workflow, we initially presented participants with a YouTube video, followed by their interaction with \systemname for real-time in-situ assistance during cooking.

We chose the recipe for a chicken mango avocado salad for our study due to its relatively short preparation time, with all the steps typically completed within 30 minutes. However, this recipe presents a cognitive challenge for users because of its numerous ingredient measurements, often necessitating external assistance~\cite{logie2010multitasking, kosch2019digital, weber2023designing}. Furthermore, to address safety concerns, the recipe does not require the use of an oven, stove, or sharp knife (instead, a table knife is used), ensuring the ethical compliance of our study.

\subsection{Recruitment Process} 
\label{sec:recruitment_process}
Participants for this study were recruited via social media platforms and email. Recruitment posters were shared along with a comprehensive description of our research objectives, a direct link and QR code leading to the screening questionnaire. The screening questionnaire was used for participant selection and included questions related to demographic information, allergy history, prior usage of CAs, and participants' cooking experiences.

A total of 12 participants were successfully enrolled in our study, each meeting the following eligibility criteria: being 18 years of age or older, fluent in English, possessing prior cooking experience, comfortable with audiovisual recording during the experiment, and having no known food allergies to the ingredients used in the study. The experimental session's duration was less than one hour. Each participant was compensated with a \$30 Amazon e-gift card for acknowledging and contributing their time to participate in our study.

\begin{table}[t!]
\centering
\resizebox{\textwidth}{!}{%
\begin{tabular}{cccccccc}
\hline
\hline
\textbf{ID} & \textbf{Gender} & \textbf{Age}
& \begin{tabular}{@{}c@{}}\textbf{Cooking}\\\textbf{Frequency}\end{tabular}
& \begin{tabular}{@{}c@{}}\textbf{Recipe}\\\textbf{Searching}\\\textbf{Frequency}\end{tabular}
& \begin{tabular}{@{}c@{}}\textbf{CA}\\\textbf{Frequency}\end{tabular}
& \begin{tabular}{@{}c@{}}\textbf{CA}\\\textbf{Recipe}\\\textbf{Search}\\\textbf{Frequency}\end{tabular}
& \textbf{CA Usage} \\ \hline
1  & Female & 18-24 & $\geq$1/week & $\geq$1/week & $\geq$1/month & Rarely & Check weather, Look up information, Small talk, Set alarm \\ \hline
2  & Female & 25-34 & $\geq$1/week & Rarely & Rarely & Never & Check weather, Home device control, Music, Set alarm, Check time \\ \hline
3  & Male   & 25-34 & Daily     & $\geq$1/week & $\geq$1/week & Rarely  & Check weather, Home device control, Set alarm \\ \hline
4  & Male   & 18-24 & Daily     & $\geq$1/week & $\geq$1/month & Rarely & Check weather, Music \\ \hline
5  & Male   & 25-34 & Daily     & $\geq$1/week & Daily & Rarely & Check weather, Music, Look up information, Set alarm, Check time, Reminder \\ \hline
6  & Female & 25-34 & Daily     & $\geq$1/week & $\geq$1/week & Rarely & Check weather, Music, Look up information \\ \hline
7  & Male   & 18-24 & Daily     & Daily & Daily  & Rarely & Check weather, Music, Look up information, Set alarm, Reminder \\ \hline
8  & Female & 25-34 & $\geq$1/week& $\geq$1/week& $\geq$1/week & Never  & Check weather, Music, Set alarm, Check time \\ \hline
9  & Female & 25-34 & Daily     & Rarely    & Rarely & Rarely & Music, Look up information \\ \hline
10 & Male   & 25-34 & $>$1x/wk  & $\geq$1/week    & Rarely & Rarely & Music, Set alarm, Reminder \\ \hline
11 & Male   & 18-24 & Daily     & $\geq$1/month  & Daily  & Never  & Check weather, Music, Reminder \\ \hline
12 & Male   & 25-34 & Daily     & $\geq$1/week  & Daily  & Never  & Home device control, Music, Set alarm \\ 
\hline
\hline
\end{tabular}
}
\caption{Overview of participant demographics and CA usage for our 12 participants. This table summarizes user habits related to cooking, recipe searches, and CA usage. Frequencies are categorized as Daily, $\geq$1/week (at least once per week), $\geq$1/month (at least once per month), Rarely, and Never.}
\label{table:demographics}
\end{table}

\subsection{Participant Demographic Information}
\label{sec:demographic}
We recruited a total of 12 participants. The sample consisted of 58.3\% male and 41.7\% female participants, majority aged between 25 and 34. Most participants reported cooking daily and searching for recipes at least once per week. CA usage patterns varied, with 33.3\% using it daily and others less frequently. Common CA activities included listening to music (83.3\%), checking the weather (75\%), and setting alarms (66.7\%), while small talk was rare (8.3\%). Recipe searches were primarily conducted on YouTube and recipe-sharing websites, with limited use of CA for cooking assistance. Detailed demographic information is available in Table~\ref{table:demographics}.

\subsection{Data Collection}
\label{sec:method-data_collection}

\subsubsection{Semi-Structured Interview}
We designed and conducted a semi-structured interview after participants finished with their post-study questionnaire. The interview covered simple questions on participants' experience using existing CAs and our system \systemname, cooking habits, and how they envision using \systemname in the future. Each interview lasted between 15 - 40 minutes.
These semi-structured interviews provided information for researchers to understand participants' overall experiences with CAs, particularly in the cooking task.

\subsubsection{Survey Measures}
In this study,  we utilize different methods to collect results from interaction, performance, subjective workload, and participants' feedback to explore our RQ1. 
A pre-study questionnaire was provided to collect participants' context on basic demographics, cooking background, and usage of voice assistants. Participants were also requested to complete a five-question survey that we designed to assess their perceptions of the CAs' capabilities.

The post-study questionnaires consisted of four elements: the Voice Usability Scale (VUS), a 12-question scale that assesses the usability of the voice interface ~\cite{zwakman2021usability}; the Explainable AI survey (XAI), a six-question scale that evaluates the trustworthiness of explainable AI systems' output from users ~\cite{hoffman2018metrics}; the NASA-Task Load Index (NASA-TLX), a six-question scale used to measure participants' subjective workload in six dimensions ~\cite{hart2006nasa}. Participants were also asked to complete the same survey provided before the study to evaluate any potential changes in their perceptions of the current capabilities of the voice assistant. The Results section will provide a detailed analysis of the survey results.

\subsection{Data Analysis Process}
\label{sec:data_analysis}
The collected survey responses were based on a 5-point Likert scale. Mean and standard deviation were calculated for each question to summarize user responses, providing a general sense of performance. Higher scores indicated better performance or more positive user experiences. The quantitative data served as a supplement to the qualitative findings, offering context to the results derived from thematic analysis.

A total of 4 hours and 35 minutes of interview audio, along with 3 hours and 39 minutes of audiovisual recordings from the experiment were collected. These recordings were transcribed using an automated service for further analysis. The co-authors independently conducted open coding for the first two participants, employing thematic analysis to identify initial themes related to the research questions.

After initial coding, the co-authors collaboratively reviewed, discussed, and categorized the codes to establish a preliminary codebook. This codebook was refined through iterative discussions and reviews of emerging codes, ensuring consistency and agreement across interpretations. The final codebook was collaboratively refined and finalized after achieving agreement among the co-authors. One author then applied the finalized codebook to the remaining transcripts following Grounded Theory~\cite{braun2006using,glaser2017discovery}.


\section{Results}
In addressing RQ1, this section presents quantitative results to provide an objective understanding of user experience. We then discuss the themes of users' experience, including successful and challenging experiences with LLM-based CAs in cooking tasks.

\subsection{Quantitative Results}
To better understand user experiences with our system, \systemname, we evaluated system accuracy through conversations (Section \ref{system_accuracy}), analyzed users’ task performance (Section \ref{task_performance}), and collected quantitative feedback through questionnaires (Section \ref{survey_result}). Although the sample size was small, the results provide a snapshot of user impressions, offering measurable insights to complement our qualitative observations.

\subsubsection{System Accuracy Evaluation}
\label{system_accuracy}
Evaluating the accuracy of \systemname in interpreting and responding to instructions is crucial. In our analysis, a response was considered valid when it contained accurate content verifiable by the recipe and was expressed fluently.Among 447 queries, 66.4\% of queries received valid and accurate responses. 23.0\% were invalid due to Alexa system errors and speech-to-text inaccuracies, while 10.6\% were invalid due to LLM errors of incorrect sequencing or unrelated answers. Participants posed a total of 28 queries beyond the recipe's scope during the study. 75.0\% of those queries received valid responses. Some invalid responses were due to the VA's inability to access specific information, such as the task's remaining time, and difficulty verifying the accuracy, such as the calorie count of the dish not being mentioned in the YouTube video or recipe.

Our accuracy calculation represents the maximum potential inaccuracy of our system, as any response that could not be verified against the original video or recipe was classified as invalid. Despite this, we observed that users often rephrased or repeated their questions to obtain valid answers, suggesting that initial invalid responses did not prevent them from continuing their tasks. Additionally, we acknowledge the inherent limitations of LLMs, such as the propensity to generate hallucinated responses, which will be discussed further in the limitations section.

\subsubsection{User Task Performance Result}
\label{task_performance}
To explore \systemname's efficacy, we analyzed key metrics such as task completion rates and performance efficiency among participants. All participants completed the assigned task within the 30-minute limit. The completion times ranged from approximately 13 minutes 19 seconds to 26 minutes 10 seconds (mean = 18 minutes 26 seconds). The average number of queries per participant was 37.7 queries, ranging from 19 to 75 questions.

Out of the 12 participants, all participants successfully prepared the dish accurately with all correct ingredients, relying on \systemname without direct access to the recipe or video while cooking. Notably, three participants precisely followed the procedures outlined in the video and instructions. As for the remaining nine participants, they introduced some sequencing errors. However, these mistakes did not impact the final dish's outcome. These deviations included multitasking and improvisation, like rearranging the order of adding garlic, sea salt, and black pepper after receiving the complete instructions (P1,2,4,6,8,10,12). 
Overall, every participant successfully completed the dish. In the following section, we will delve into their experiences interacting with the system by analyzing the survey results.

\subsubsection{Survey Result}
\label{survey_result}
We employed four scales in this study: the Voice Usability Scale (VUS) to evaluate the usability, affectiveness, and recognizability \& visibility of the voice interface~\cite{zwakman2021usability}, the XAI survey to assess the trustworthiness of system outputs ~\cite{hoffman2018metrics}, and the NASA-TLX to measure task workload, including mental, physical, and temporal demands, effort, frustration, and performance ~\cite{hart1988development}. These surveys were adapted to fit \systemname by selecting relevant questions while omitting those unrelated to the experimental task. The complete set of questions and detailed results are provided in Appendix~\ref{appendix:questionnaire}, and responses were collected using a 5-point Likert scale, where higher scores indicated better performance.

Overall, participants perceived \systemname positively in cooking scenarios. The VUS results reflected overall satisfaction with the system’s user experience, evaluating usability, affectiveness, and recognizability \& visibility. The usability questions assessed the system's difficulty, resulting in a score of 2.03 on a 5-point Likert scale. For affectiveness, participants rated the system with a mean score of 4.25 on a 5-point Likert scale. Finally, the mean score for recognizability and visibility was 3.25 on a 5-point Likert scale.

The XAI survey measured trustworthiness across four dimensions: predictability, reliability, efficiency, and believability~\cite{zwakman2021usability}. Results for all questions exceeded a mean score of 3.9, with an overall trustworthiness score of 4.11 on a 5-point Likert scale. NASA-TLX results showed low levels of mental, physical, and temporal demand, as well as low frustration, alongside high-performance ratings. These quantitative findings align with qualitative interview feedback, where most participants described successful interactions with minor obstacles.

Beyond these scales, we explored whether participants’ perceptions of CAs changed after interacting with \systemname. We included five exploratory questions asked both before and after the study. These questions focused on conversational fluency, memory, follow-up questions, integration into daily activities, and active collaboration. Results across all five aspects improved after participants used \systemname.

\subsection{Themes of User's Experience with LLM-Based CAs}
To further understand users’ perceptions through their interaction experiences, we categorized users’ experiences into several themes through thematic analysis.
In the following sections, we will present these themes under two high-level categories: successful and unsatisfactory, based on the interaction between users and \systemname. Successful experiences were those where users effectively leveraged the LLM’s capabilities, receiving clear, actionable instructions that enhanced their cooking tasks and met or exceeded their expectations. In contrast, unsatisfactory experiences occurred when users encountered difficulties in utilizing the capabilities of LLM during their cooking tasks.

\subsubsection{\textbf{Successful} Experience When Using LLM-Based CAs for Cooking Tasks} \label{successful_exp}\hfill \break
From the survey results, we have confirmed that participants had an overall successful experience using \systemname. In this section, we will elaborate on specific aspects of participants' usage and experiences they were satisfied with, particularly those related to the LLM powered capabilities.

Firstly, many participants asked \systemname for \textbf{information that extended beyond the scope of the recipe and received satisfactory answers}. These inquiries often revolved around fundamental cooking tips, which might be unrelated to the specific recipe and were not included in the original instructions. 
These were particularly helpful, especially for novice cooks lacking essential cooking knowledge. 
For instance, P5 inquired, ``How do you peel an avocado?'' Such information not only aided in the immediate task but also contributed to participants' overall cooking skills. 
Another category of information sought by users pertained to the recipe but was not explicitly provided in the original instructions, such as nutrition information. For example, P8 asked, ``How many calories are in the salad?'' \systemname responded with an estimate: ``This might answer your question, 224 calories,'' despite the absence of this specific data in the original recipe. 
Despite the lack of explicit information, ChatGPT is capable of estimating the results and providing a suggestion. 
Importantly, participants posed these questions naturally, demonstrating their recognition of the system's ability to address such inquiries.

\begingroup
\renewcommand\arraystretch{1.2}
\begin{table}[b!]
\begin{scriptsize}
  \centering
  \begin{tabular*}{0.97\linewidth}{@{\extracolsep{\fill}}l l l}
    \toprule\toprule
    \makecell[cl]{\textbf{Theme}}
    & \makecell[cl]{\textbf{Sub-theme}}
    & \makecell[cl]{\textbf{Example}}\\
    \midrule
    \makecell[tl]{\textbf{Receive Extensive }\\ \textbf{Information Beyond} \\ \textbf{the Recipe by the LLM}}
    & \makecell[tl]{Fundamental Cooking Tips}    
    & \makecell[tl]{``Alexa, how to peel avocado?''(P5) \\
    ``Alexa, tell me that amount of teaspoon if I want one quarter \\
    tablespoon''(P1)}\\
    & \makecell[tl]{Nutrition Information \\ Related to The Dish}
    & \makecell[tl]{
    ``I asked how many calories are in the in the salad'' (P8)}\\
    \midrule

    \makecell[tl]{\textbf{Contextual Memory}\\ \textbf{\& Task Awareness}}
    & \makecell[tl]{Current Step}    
    & \makecell[tl]{``There was one question I asked which step am I at right now? \\And he told me on step four.'' (P5)}\\
    \midrule

    \makecell[tl]{\textbf{Adaptive Contextual}\\ \textbf{Personalization}}
    & \makecell[tl]{Lack of Tools}    
    & \makecell[tl]{``Alexa, I want to I want to make the lettuce dry without some \\water but I don’t have a spinner so how can I do it?'' (P1)}\\
    \midrule

    \makecell[tl]{\textbf{Plan Tasks \& Control Flow }\\ \textbf{Dynamically}}
    & \makecell[tl]{Support Multi-Tasking/ \\ Task Planning}    
    & \makecell[tl]{``I always used to ask it a few steps before. So when I'm cutting\\ the onion, I would ask what I need to do with a tomato. '' (P10)\\
    ``When I focus on something I just asked, you know, \systemname \\whats the next step and then I was cutting stuff and it says the \\instruction.''(P4)}\\
    & \makecell[tl]{Change the Order of Tasks}    
    & \makecell[tl]{``And basically, I could execute things in my order as well. I did \\ not have to follow the same path, I could figure out my own \\ path.'' (P10)}\\
    \midrule

    \makecell[tl]{\textbf{Culinary Learning}\\ \textbf{through Recipes}}
    & \makecell[tl]{New Recipes}    
    & \makecell[tl]{``I would say it works well on beginners and people who have like\\ a good experience with cooking but who are also new to certain\\ recipes.''(P10)}\\
    \midrule
    
    \makecell[tl]{\textbf{Conversational Engage-}\\ \textbf{ment and Encouragement}}
    & \makecell[tl]{Congratulation Messages}    
    & \makecell[tl]{Q: Do you think this Alexa talks differently?\\ P2: Expressions like enjoy your food.}\\
    \bottomrule
    \bottomrule
  \end{tabular*}
  \caption{Qualitative code book and description of participants' \textbf{successful experience} and usage with the LLM-based CA.}
  \label{tab:table_pros}
\end{scriptsize}
\end{table}
\endgroup

Another common type of question participants frequently asked was \textbf{next-step instructions}, such as ``What's the current step?'' This pattern of inquiry suggests that users had recognized the system's capability to remember the ongoing status and the history of the conversation. 
P6 pointed out, ``It follows up on your previous question… It sticks to the track, so it's like one continuous flow.'' Similarly, P1 was impressed by the system's ability to stay on track, stating, ``It (\systemname) can memorize which step you are in right now. And you can continue to the next one instead of starting over from the very beginning.'' When interacting with the system, users quickly accepted the fact that \systemname could retain this information, indicating a high level of confidence in its retrieval capabilities as a 'machine'.

\systemname also excelled in \textbf{tailoring solutions to meet users' specific requirements}, and our participants quickly took advantage of this feature to receive instructions based on their own settings. For instance, during the experiment, some participants encountered challenges when a specific tool demonstrated in the video or recipe was unavailable. In these situations, our system provided valuable assistance, even when these occurrences were not explicitly outlined in the original instructions given to our system. P1 noticed a missing tool and asked for help from \systemname, stating, ``Alexa, I want to make the lettuce dry without a spinner, but I don't have one. How can I do it?'' \systemname offered tailored, step-by-step guidance on completing the task without the missing tool. 
Moreover, \systemname's responses could be further personalized based on the specific setting and individual user preferences. In a different instance during the experiment, P10 asked, ``Alexa, give me all the vegetables and leafy greens that I need to chop,'' Instead of following the procedure described in the original recipe, \systemname responded with customized instructions: ``You will need to chop one-quarter head of romaine lettuce, English cucumber, and thinly sliced purple onions for the salad.'' This shows that our participants had both needs and confidence in \systemname's ability to reorganize existing information to tailor it to users' needs.

Due to its extensive capability to customize instructions according to users' requests, participants also realized its ability to assist them in \textbf{planning cooking tasks and dynamically controlling the workflow}. Participants could adjust the order of tasks based on real-time situations or even plan for multitasking with the assistance of \systemname. For example, P10 preferred to inquire about tasks a few steps in advance, stating, ``I always used to ask it a few steps before. So when I'm cutting the onion, I would ask what I need to do with a tomato.'' P10 also highlighted the advantages of this approach with \systemname's assistance, noting, ``I wouldn't be standing there waiting for it to give me an answer. I would always be doing something… You get to ask a question one step ahead at a time, and that helps.'' Likewise, P4 adopted a similar strategy for task planning, saying, ``When I focus on something, I just asked \systemname what the next step was, and then I was cutting stuff, and it gave me the said instruction.'' In summary, \systemname's ability to promptly react to in-situ flow changes enables more efficient and dynamic flow control for users, especially those with advanced skills in task planning within cooking scenarios. However, it is important to note that instead of providing goals and letting \systemname plan the order of tasks, our participants tended to only ask for information and still did the planning themselves. This suggests a potential preference for a usage mode in cooking, which often requires complex task planning and extensive user controls.

In addition, participants praised the system's ability to help them learn a new recipe, which was the case in the experiment where all the participants made this specific salad for the first time. Acknowledging this learning potential suggests that our participants may view \systemname as a mentor-like system with extensive knowledge of the recipe and general cooking, capable of teaching them new things they were unaware of.

Lastly, an interesting response came from P2 when we asked, ``Do you think Alexa talks differently?'' They answered, ``Expressions like `enjoy your food'.'' While this information may not be necessary for completing the task, it mimics human-like conversation and fosters a sense of conversational engagement and encouragement, making the participant feel like they are interacting with a special CA. This experience shows that having such human-like can positively influence a participant’s perception of the system, enhancing the overall user experience.

In summary, we explored participants' successful experiences and interactions with \systemname. In the following section, we will delve into some of the unsatisfactory experiences. 

\subsubsection{\textbf{Unsatisfactory} Experience When Using LLM-Based CAs for Cooking Tasks}\hfill \break
\begingroup
\renewcommand\arraystretch{1.2}
\begin{table}[b]
\begin{scriptsize}
  \centering

  \begin{tabular*}{\linewidth}{@{\extracolsep{\fill}}l l l}
    \toprule
    \toprule
    \makecell[cl]{\textbf{Theme}}
    & \makecell[cl]{\textbf{Sub-theme}}
    & \makecell[cl]{\textbf{Example}}\\
    \midrule
    \makecell[tl]{\textbf{Information Overload}}
    & \makecell[tl]{Too Many Ingredients \\at Once}    
    & \makecell[tl]{``The first was that he was giving too much information. Like for example, he's\\
    telling me salt and pepper together where I have to measure one, measure the \\
    other one where we measure the first one. And I forgot about that.'' (P5)}\\
    & \makecell[tl]{Speak Too Fast}
    & \makecell[tl]{``When \systemname actually provide me with the steps it's kind of speak too \\
    fast and I have to kept asking \systemname to repeat the instructions''(P8)}\\
    \midrule

    \makecell[tl]{\textbf{Misunderstanding of}\\ \textbf{Oral Expressions}}
    & \makecell[tl]{Linguistic Error for\\ Oral Expressions}    
    & \makecell[tl]{``One area where I found a mistake was that I asked what’s the `last' instruction,\\ 
    meaning the `previous' one, it took me to the `very last' instruction.''(P7)}\\
    \midrule

    \makecell[tl]{\textbf{Increased Cognitive} \\ \textbf{Load}}
    & \makecell[tl]{Distracting Back and \\ Forth Conversation}    
    & \makecell[tl]{``Even though it's doing a very good job with interaction, sometimes you still\\
    need to try talk to it slowly so you can understand the answers. That requires like\\
    back and forth conversation. But while you were doing that, and if something is\\
    on the stove, that could be very distracting.''(P4)}\\
    \midrule

    \makecell[tl]{\textbf{Expect More Dialogue} \\ \textbf{with the System}}
    & \makecell[tl]{Lack of Verification \\ From Users} 
    & \makecell[tl]{``I would like it to repeat my questions. So that I understand that I'm giving the \\
    right instruction. It's much better than I asked something, and it (\textit{Mango Mango}) \\ 
    misunderstand me and gave the wrong answer.''(P8)} \\
    && \makecell[tl]{``I would like to show me more itself or asked me for confirmation in my case.'' (P8)} \\
    \midrule

    \makecell[tl]{\textbf{Only Passive Response}}
    & \makecell[tl]{Lack of Auto-Tracking}    
    & \makecell[tl]{``I would like a mode in which, for example, while making the dressing, instead of\\
    telling me all the ingredients one after the other and may not be able to catch up.\\
    Maybe if I asked her (\systemname) to like check on me before proceeding. That \\
    would be like an amazing step, amazing feature to have.'' (P10)}\\
    \midrule

    \makecell[tl]{\textbf{Feature Discovery}\\ \textbf{Challenges}}
    & \makecell[tl]{Lack of User Guidance \\on System's Features}    
    & \makecell[tl]{``I don't know whether Alexa can help me to control the time or it can just tell me.\\ I don't know like whether it can intelligently tell
    me when I should maybe do\\ something and do the other things. I don't know
    whether he (\systemname) can\\ do that.'' (P9)}\\
    \bottomrule
    \bottomrule
  \end{tabular*}
  \caption{Qualitative code book and description of participants' \textbf{unsatisfactory experience} and usage with the LLM-based CA.}
  \label{tab:table_cons}
  \end{scriptsize}
\end{table}
\endgroup
Although our participants benefited greatly from the assistance provided by \systemname, there were still many challenges during the interaction, many of which related to the disparities of perception in the LLM powered capability.

There are instances of dissatisfaction from users due to \textbf{information overload}. In our experiment, the recipe encompassed instructions for preparing the salad and crafting the dressing. Although all the necessary ingredients were provided, participants were tasked with measuring precise amounts for the dressing. The dressing was introduced all at once in the instructional video, and we followed a similar approach in our written recipe prompt. However, many participants expressed difficulties following \systemname’s instructions, primarily due to the presentation of multiple ingredients at once. For instance, P5 articulated this issue, stating, ``The first was that it was giving too much information. For example, he’s telling me salt and pepper together, where I have to measure one and then measure the other one. But when I measured the first one, I forgot about the other one'' Additionally, P8 also highlighted the narrative speed was too fast, ``When \systemname delivers the instructions, it tends to speak too rapidly, necessitating repeated requests for clarification'' This indicates the system's inability to comprehend and deliver the appropriate amount of information, which, however, is a fundamental requirement for users to ensure fluent and informative conversation.

Furthermore, as users became more accustomed to natural conversations with \systemname, some \textbf{system constraints} became more evident, such as misunderstandings of oral expressions, the need to initiate conversations using the wake word, and increased cognitive load. For example, P7 encountered a linguistic error during the experiment and noted, ``One area where I found a mistake was that I asked what’s the `last' instruction, meaning the `previous' one, it took me to the `very last' instruction.'' This occurred due to the ambiguity of certain words, which can have multiple meanings, especially in oral versus written contexts. 
Additionally, when users are able to adopt the system's assistance for complex tasks like multitasking, it introduces a significant cognitive load compared to simply listening to instructions. In the case of cooking, this increased cognitive load could potentially hinder task completion and even lead to safety issues. In summary, we observed that as our conversations became more natural due to the extensive capabilities provided by the LLM, the system needed to adapt accordingly. It had to recognize that the dialogue had become more oral, making it more challenging for users to consistently use the wake word. Additionally, the increased cognitive load needed to be addressed.

Recognizing that the system is imperfect and sometimes does not behave as expected, some participants expressed a desire for our system to incorporate more user feedback for further verification before making decisions. P8 highlighted a specific suggestion: ``I would like it (\systemname) to repeat my questions so that I can confirm that I’m providing the right instruction. It’s much better than asking something, and it (\systemname) misunderstands me and gives the wrong answer.'' To address this issue more effectively, P8 expressed a preference for the system to ``show me more itself or ask me for confirmation in my case'' to minimize the occurrence of misunderstandings. We realized that although our system supports further iteration through follow-up questions, it was primarily designed as a question-solving system that often aims to provide an immediate answer rather than engaging in cooperative decision-making with users. As an LLM-based assistant, users expect it to be more communicative and involve them more in decision-making.

Similarly, as a question-initiated system, \systemname primarily provides \textbf{passive responses}. However, P10, for instance, expressed a desire for the system to ``check on me before proceeding.'' We realized that this indicates an increased level of expectation from users. ``Checking on users'' suggests a transition from the system being a passive assistant that waits for questions to an `agent' that actively participates in the process and provides assistance. Note that P10 also mentioned ``checking on me'' rather than ``telling me what to do,'' which aligns with the earlier statement about users preferring more dialog-like suggestions rather than direct instructions.

Finally, P9 raised a notable concern regarding \textbf{the absence of clear guidance on the available features} when using \systemname, which is unsurprising. Despite the advantages offered by \systemname, participants were constrained by a 30-minute time limit for task completion, coupled with a brief 5-minute tutorial provided by the research team before initiating the assignment. P9 articulated this issue by saying, ``I don't know whether Alexa (\systemname) can help me control the time or intelligently tell me when I should maybe do something and do the other things. I don't know whether it (\systemname) can do that.'' Considering this, presenting the full range of \systemname's capabilities could potentially empower users to use it more effectively and extract maximum benefits, especially in real-world contexts. However, how to design such a tutorial remains an issue that needs further discussion, which we will also explore in a later section.

In summary, we explored participants' unsatisfactory experience and interactions with \systemname. In the following section, we will discuss how this might influence future design.

\section{Discussion}

In this study, we explored users’ successful and challenging experiences while interacting with \systemname in a cooking scenario, specifically focusing on the different interactions facilitated by the LLM. Based on the insights from the user studies, we address RQ2: What are the design implications of LLM-based CAs aimed at assisting users in real-world practices like cooking? (Section \ref{design_considerations}). Additionally, we expand on RQ1 by discussing users’ shift from traditional recipes to LLM-CAs as instructional tools for cooking (Section \ref{form_of_recipe}, \ref{llm_ca_tool}). Next, we discuss other potentials and challenges of LLM-based CAs in real-world practices inspired by our study results, presenting directions for future works. (Section \ref{privacy}, \ref{hallucination}, \ref{accessibility}). Finally, we recognize the study’s limitations (Section \ref{limitation}).

\subsection{Design Considerations: Redesign LLM-CAs for Effective Collaboration}
\label{design_considerations}
In this section, we will delve into how our findings guide the design of CAs capable of leveraging the full potential of LLM to assist users in accomplishing real-time tasks. We proposed five design implications for a future LLM-CA, offering actionable solutions and providing examples in the context of cooking.

\subsubsection{Contextualize an Universal LLM-CA for Specific Tasks}\hfill \break
In contrast to typical CA applications that rely on techniques like intent recognition in NLP~\cite{hamada2005cooking, nouri2019supporting, sato2014mimicook}, leveraging LLM provides an easier approach to handling a wide array of inquiries beyond rigid rule-based frameworks through prompts engineering~\cite{hwang2023large,wei2023leveraging,lee2023dapie,wang2023enabling}. In the past, various dialogue systems attempted to encompass as many user queries as feasible within their training resources due to limitations in understanding questions beyond those resources~\cite{mctear2018conversational,hocutt2021interrogating,vtyurina2019verse}. However, with the robust capability to comprehend 'common sense knowledge'~\cite{wei2022chain}, the challenge for an LLM-powered CA shifts to contextualizing a diverse range of inquiries. Past CAs are usually compartmentalized for specific applications. In contrast, LLM-CA often possesses the ability to manage multiple applications through a single agent, making context identification even more challenging. Although prompt engineering might offer partial solutions for this issue~\cite{lo2023clear,poola2023overcoming, wang2023prompt}, the distinct challenge of understanding these questions in real-time voice-based tasks such as cooking persists due to the inherently oral nature of these inquiries. In our experiment, we noticed that users' queries were always oral and vague, often lacking direct references to the specific recipe or the type of tasks. As a result, apart from describing the context in prompts ~\cite{liu2021makes}, to ensure smooth context transitions and maintain relevance, we propose the following solution:
\vspace{-1mm}
\begin{itemize}
    \item Extract context from conversation history to help understand users' queries. In cooking scenarios, this can involve identifying the user's current cooking step based on the past conversation and offering relevant instructions accordingly.
    \item Proactively ask for user confirmation. The system could proactively verify the user's current stage during the cooking process. This check-in ensures the system's responses align with the user's actual progress and needs.
\end{itemize}

\subsubsection{Augment Current LLM Knowledge Base}\hfill \break
LLM systems have the capability and should actively gather information to enrich their knowledge base, specifically related to the target tasks~\cite{gao2023retrieval,czekalski2024efficiently}. In our experiment, by integrating relevant information such as ingredients, steps and instructions as an external knowledge base, \systemname performed well in responding to queries about making the salad. While the current system performs sufficiently, it could further enhance its effectiveness by integrating a border range of task-specific information. During the experiment, participants raised questions that extended beyond the recipes, such as asking about the salad's calories. Broadening the system's specialized knowledge base is recommended to address such inquiries effectively, enabling users to receive accurate responses to their related questions. Consequently, we propose:
\vspace{-1mm}
\begin{itemize}
    \item Incorporate special training materials. For cooking scenario, this could involve integrating nutritional information and fundamental cooking techniques into its knowledge base.
    \item Fine-tune the model according to the context of specific tasks.
\end{itemize}

\subsubsection{Elicit Feature Discovery Instead of Focusing on the Expectation Alignment}\hfill \break
The advancements in knowledge of LLM-CA as described in the earlier section also necessitate a shift in how systems should engage with users to convey their usability. Prior studies have highlighted that voice assistants (without LLMs) often fall short of providing comprehensive knowledge, proposing the design implications for systems to focus on communicating their capabilities and limitations~\cite{hwang2023rewriting,hocutt2021interrogating}. However, this design consideration needs adaptation, as now the knowledge base of LLMs might occasionally exceed users' expectations instead of falling short. In such cases, instead of limiting users in their interactions with the voice assistant, the system should encourage diverse and creative inquiries. Therefore, LLM-CAs should proactively reveal hidden features that support dynamic, creative, and context-specific user interactions and encouraging users to explore these hidden features could increase engagement and satisfaction with the system.

A mechanism for supporting this shift from expectation alignment to feature discovery is prompt design, which defines how the LLM-CA responds to user queries and encourages user-driven exploration. While system developers design the initial prompt that establishes the assistant’s tone, role, and operational logic, users also act as ``prompt engineers'' during interactions. Unlike traditional VAs, where interactions are pre-programmed intents, LLM-CAs allow users to issue open-ended, custom prompts.
However, not all users know how to prompt effectively. Research shows that users are often unfamiliar with prompt engineering and struggle to phrase their requests in ways to get responses that they are looking for~\cite{zamfirescu2023johnny}. This knowledge gap can leave users unaware of system capabilities, ultimately limiting engagement. To bridge this gap, we propose: 
\vspace{-1mm}
\begin{itemize}
    \item For system prompt design, the system’s initial prompt should establish its role, such as a ``cooking assistant'', to offer task-relevant guidance, prevent unrelated responses, and support users in discovering relevant features.
    \item Add suggestions at the end of relevant responses to guide users on how to phrase their queries. For instance, after telling the user to add salt to the salad, the system could add a note (e.g., \textit{``You can also ask for help if you've added too much salt.''}) to teach users effective query patterns.
\end{itemize}

\subsubsection{Calibrate Trust in System Accuracy and Reliability}\hfill \break
While acknowledging the extensive capabilities of LLM-CAs, we also recognize the challenges posed by integrating LLMs into voice assistants. LLMs are known to suffer from issues such as hallucination caused by overconfidence in responses, sycophancy and more~\cite{kamath2024llm}. These challenges can potentially decrease user trust and result in harmful outcomes.

It is still important for humans not to fully rely on LLM-CAs, as humans can apply prior knowledge and common sense to validate responses~\cite{shankar2024validates,abeysinghe2024challenges}. This becomes particularly critical when user input diverges from the LLM’s responses, which may lead to hallucination, particularly when speech-to-text errors occur~\cite{koenecke2024careless}. In our study, participants were shown a recipe video before using the system, giving them a baseline understanding of the steps expected from the LLM-CA. However, our system occasionally generated unreliable answers, though these instances were not frequent. One notable example occurred when P1 mentioned having only a quarter cup of olive oil, while the recipe required 1/8 cup. Instead of advising how to measure the required amount, \systemname incorrectly suggested altering the recipe to accommodate a quarter cup.

To enhance the collaborative experience, it is important to balance encouraging user exploration with effectively communicating the reliability of answers to users, thereby building a trustworthy system ~\cite{shen2023chatgpt, raj2023semantic, cuadra2022inclusion}. Maintaining transparency about the system’s capabilities can familiarize users with LLM-CAs strengths and limitations, fostering more collaborative interactions ~\cite{liao2023ai,zhang2024sa,zhang2024can}. As detailed in Sec.~\ref{system_accuracy}, these types of misguidance highlight the importance of careful design considerations when creating LLM-based voice assistants for complex tasks. Therefore we propose the following recommendations to mitigate such issues in LLM-CAs:
\vspace{-1mm}
\begin{itemize}
    \item Indicate the level of confidence in responses. In cooking scenarios, after providing an uncertain answer, the system could add a statement like, \textit{“I am not very sure about this response since it's outside of the recipe,”}. This will help build user's trust in the system's reliability.
    \item Including brief reasoning in responses. This helps users actively engage with the system, especially in tasks where accuracy matters or trust is important. It allows users to quickly check if a response is correct and provide feedback or ask for clarification, supporting shared decision-making. In critical situations where mistakes could have serious consequences, such as cooking with food on the stove, involving users in validating the response builds trust and creates a stronger collaboration between the user and the system.
\end{itemize}

\begin{table}[t]
\begingroup
\renewcommand\arraystretch{1.2}
\begin{scriptsize}
\begin{longtable}{llll}
\toprule
\toprule
\centering
\textbf{Design consideration} &
\textbf{Example from experiment} &
\textbf{Solution}& \textbf{Example in cooking}\\ \hline
\endfirsthead
\endhead
\begin{tabular}[t]{@{}l@{}}
\textbf{1. Contextualize a}\\ \textbf{universal LLM-CA}\\ \textbf{for specific tasks}\end{tabular}
&
    \begin{tabular}[t]{@{}l@{}}
    Many inquiries are lack of direct\\references to the recipe and the\\type of tasks.
    \end{tabular} 
&
    \begin{tabular}[t]{@{}l@{}}\textbf{Extract context from} \\\textbf{conversation history}\\\textbf{Proactively ask for} \\\textbf{user confirmation.} \end{tabular} 
&   
    \begin{tabular}[t]{@{}l@{}}Offer step guidance based on\\conversation history\\Proactively verify user's current\\cooking stage.
    \end{tabular} \\
\midrule
\begin{tabular}[t]{@{}l@{}}
\textbf{2. Augment current}\\ \textbf{LLM's knowledge}\\ \textbf{base}\end{tabular} 
&
    \begin{tabular}[t]{@{}l@{}}
    Question like \textit{``How many calories}\\ \textit{are in the salad?''(P8)} could not \\be answered accurately using\\ current knowledge base.
    \end{tabular} 
&
    \begin{tabular}[t]{@{}l@{}}
    \textbf{Incorporate special} \\\textbf{training materials}\\ \textbf{Fine-tune the model}\\ \textbf{on specific task.}
    \end{tabular} 
&
    \begin{tabular}[t]{@{}l@{}}
    Incorporate information such as\\nutritional details and fundamental\\ cooking techniques.
    \end{tabular}\\
\midrule
\begin{tabular}[t]{@{}l@{}}
\textbf{3. Elicit feature}\\ \textbf{discovery instead}\\ \textbf{of expectation}\\ \textbf{alignment}\end{tabular} 
&
    \begin{tabular}[t]{@{}l@{}}
    \systemname can provide\\ guidance on correcting cooking\\ mistakes, but no users are aware \\of this feature. \end{tabular}
&
    \begin{tabular}[t]{@{}l@{}}
    \textbf{Add suggestions}\\\textbf{at the end of relevant}\\\textbf{response.}
    \end{tabular} 
&
    \begin{tabular}[t]{@{}l@{}}
    \textit{System:``..., I can also help if you've}\\\textit{added too much salt.''}
    \end{tabular}\\
\midrule
\begin{tabular}[t]{@{}l@{}}
\textbf{4. Calibrate trust}\\\textbf{in system accuracy}\\\textbf{and reliability}\end{tabular} 
&
    \begin{tabular}[t]{@{}l@{}}
    \textit{P1: ``I've only got a quarter cup.''}\\
    \textit{MM: ``In that case, you can use }\\\textit{one quarter cup of extra virgin }\\\textit{olive oil instead of 1/8 cup.''}
    \end{tabular}
&
    \begin{tabular}[t]{@{}l@{}}
    \textbf{Indicate the level of}\\\textbf{confidence in}\\\textbf{responses.}
    \end{tabular} 
&
    \begin{tabular}[t]{@{}l@{}}
    \textit{System:``..., I am not very sure about} \\\textit{the response since it's outside of the} \\\textit{recipe.''}
    \end{tabular}\\
\midrule
\begin{tabular}[t]{@{}l@{}}
\textbf{5. Implement an}\\ \textbf{adaptive response}\\ \textbf{style} \end{tabular} 
&
    \begin{tabular}[t]{@{}l@{}}
    Trade-off between offering a \\ list of ingredients in response \\and not enough details on\\ how to complete a step. \end{tabular} 
&
    \begin{tabular}[t]{@{}l@{}}
    \textbf{Rephrase the}\\\textbf{response based on}\\\textbf{implicit expressions}\\\textbf{in user's request }\\\textbf{Respond as user's}\\\textbf{perception of the }\\\textbf{system role.}
    \end{tabular}
&
    \begin{tabular}[t]{@{}l@{}}
    \textit{User: ``What are the ingredients again?''}\\\textit{System:``Let me repeat with less }\\\textit{information...''} \\System as tool, personal assistant or\\partner during cooking.
    \end{tabular}\\
\bottomrule
\bottomrule
\cr\caption{Summary of the five design considerations with interaction examples from our experiment, corresponding solutions, and example solutions in the context of cooking.}
\label{tab:design_consideration}
\end{longtable}
\end{scriptsize}
\endgroup
\end{table}

\subsubsection{Implement an Adaptive Response Style}\hfill \break
Traditional CAs, designed for simple tasks like setting timers or playing music, operated within rigid rule-based frameworks and limited question sets. Previous research recommended concise, straightforward responses for transactional interactions where flexibility or deeper engagement was unnecessary~\cite{haas2022keep}. However, further studies on using traditional CAs for tasks like cooking revealed key challenges. For example, in cooking scenarios, prior investigations also revealed tradeoffs between information overload and insufficient detail in responses~\cite{hwang2023rewriting,corbett2016can,kim2021designers}.

While \systemname utilized LLM to partially address the challenge by providing adaptive responses, the unique capabilities of LLMs also offered opportunities to re-evaluate and improve design implications. One significant opportunity is the ability to address the challenge of information overload better when working with CAs, especially in cooking scenarios~\cite{hwang2023rewriting}. Previous research suggested that to mitigate information overload, assistants should carefully manage how they process and deliver information~\cite{hwang2023rewriting,jaber2024cooking}. With LLMs, there is now greater flexibility to dynamically tailor responses based on the user questions with prompting instructions, including customizing both the content and tone of responses. In our system design, we incorporated prompts instructing the system to answer questions by following the rule with fewer words and providing a response depending on the user's question. For example, the system should include the measurement details only when users ask about ingredient measurements. However, if users ask for required ingredients, the system should exclude measurement details to avoid unnecessary cognitive load. Similarly, the system adapts its tone based on the conversational context. For instance, in casual, low-urgency interactions, a humorous tone can enhance user engagement and learning~\cite{hwang2023rewriting}. In contrast, for more urgent or task-critical contexts, a clear, direct tone is preferred. However, some users in our study reported that they forgot the other ingredients mentioned after adding one ingredient. Therefore, the conversational context in complex tasks is also important~\cite{jaber2024cooking}. Unlike traditional CAs, LLM-CAs in our study leverage conversation history to deliver more contextual and relevant responses. Using LLMs, we suggest an opportunity for greater flexibility in tailoring responses based on the task context. Tasks with low cognitive load, such as placing items in a bowl, can be grouped with related steps to enhance efficiency. Conversely, high cognitive tasks, like measuring ingredients, should be presented individually to minimize the risk of overwhelming the user. 

In sum, with the capabilities of LLMs, we’ve introduced a novel response style—adaptive—that dynamically adjusts the tone and amount of information in responses throughout the duration of a task by integrating user feedback. We suggest three methods for implementing the adaptive response style:

\vspace{-1mm}
\begin{itemize}
    \item Rephrase the response through implicit expressions in user's request. Although not always explicitly stated as a command, the expressions in a user's request often reflect whether and how the future response style should be adjusted. For instance, when a user asks the system to repeat the necessary ingredients, it often implies that the previous response from the system contained too much information, prompting the need for a more concise instruction next time.
    \item Respond as user's perception of the system role. Through our analysis, we discovered that users perceived our system's role differently in their interactions. As a result, in real-time scenarios, the system needs to discern these roles and stages based on user responses. In our cooking experiment, we identified three roles (tool, personal assistant, partner) that should dictate tailored responses.
    \item LLM-CAs can be prompted to recognize the urgency and complexity of a situation based on contextual cues, such as when a user is handling tasks that require immediate attention. Instead of relying solely on the length of responses, the system should adjust the response style according to task completion metrics like task complexity, cognitive load, and urgency. For instance, even a brief response like an ingredient name can impose a high cognitive load, as users need to locate, measure, and add ingredients. In such cases, the system should avoid combining multiple instructions.
\end{itemize}

\subsection{The Shift From Other Forms of Recipes to LLM-CAs}
\label{form_of_recipe}
In our study, participants began by watching a YouTube version of the recipe before using the LLM-CA to prepare the salad. Watching the video gave participants a visual reference of the recipe’s general flow, ingredients, and key techniques. However, many participants noted that they often needed to rewatch parts of the video while following the instructions when cooking with new recipes, especially for the ingredient measurements. In contrast, with \systemname, participants mentioned they no longer needed to refer back to the video repeatedly. Instead, they could receive step-by-step guidance directly from the LLM-CA, allowing them to maintain their cooking flow.

LLM-CAs differ from traditional recipe formats, such as YouTube videos, text-based instructions, and human guidance, in supporting users during cooking tasks. One notable shift participants highlighted was the ability to engage with the LLM-CA for on-demand, context-specific support. Unlike static video instructions, which require users to pause, rewind, or search for specific information, \systemname allowed users to ask questions during the cooking process without stopping. For instance, participants mentioned that if they had questions mid-cooking—like how many calories are in the salad or clarification about a step—they could directly ask the LLM-CA. This interaction level contrasts with searching for answers online, typically requiring stopping the cooking process, navigating a device, and sifting through search results. Furthermore, P5 mentioned perceiving LLM-CAs as more ``teacher-like'' than traditional recipe formats. Instead of passively following instructions, participants felt that \systemname acted as a supportive instructor, offering contextual guidance and adapting to user input. This is distinct from following a static text or video recipe, where the user must interpret instructions independently.

Despite these benefits, participants noted that LLM-CAs cannot fully replace human guidance. Both P7 and P12 mentioned that in situations where their mother is unreachable, they would be more inclined to turn to \systemname for guidance, especially when cooking from scratch. LLM-CAs serve as a secondary source of assurance, helping users validate their cooking process. However, they are not a full replacement for the experience of seeking advice from a mother, as human connection and expertise provide different connections that LLM-CAs cannot fully replicate.

\subsection{LLM-CAs as Motivational and Instructional Tools in Cooking}
\label{llm_ca_tool}
In our study, we envisioned \systemname as both an instructional tool, guiding users through completing their cooking tasks.
However, our analysis also revealed that \systemname could serve as a motivational tool, promoting a learning procedure while completing the cooking tasks.
This finding aligns with prior research showing that CAs can support user learning by providing tailored guidance and fostering engagement~\cite{kasneci2023chatgpt}. With LLM-CAs, our participants envision learning customized content in new and meaningful ways. For example, one participant expressed a desire to use \systemname to preserve their mother’s unique recipes and revisit them anytime. Its ability to adapt to user-specific content enables self-directed learning experiences~\cite{lin2024exploring, dizon2024chatgpt}. By tailoring responses to individual needs, \systemname can serve as a repository for culturally significant or family-specific recipes, creating a bridge between technology and personal heritage.

Participants also highlighted the potential of LLM-CAs to motivate and reinforce learning through interactive and engaging responses. For example, integrating features such as real-time feedback on user progress could enhance skill development during tasks like cooking. By actively engaging with users through tailored guidance and motivational cues, LLM-CAs could transform routine activities into opportunities for experiential learning. This ability to combine instructional guidance with motivational support underscores the unique potential of LLM-CAs to enhance learning in real-world contexts.

\subsection{Integrating Multi-Modal Features and Addressing Privacy Concerns}
\label{privacy}
In our study, participants highlighted the potential of integrating multi-modal features, such as a camera, to enhance the capabilities of \systemname. By incorporating real-time visual recognition from a camera, the LLM-CA can use this visual information as additional context to acknowledge the current step or status of the food, enabling the LLM-CA to provide more accurate and context-aware guidance. Advances in multi-modal LLMs, such as GPT-4V, offer a promising way to realize this potential. With its ability to process both visual and textual inputs, facilitating more collaborative, adaptive, and context-aware systems~\cite{wang2024large}.
For example, in cooking scenarios, a camera could allow the system to identify whether the ingredients were placed or the progress of a recipe step. By processing visual input as contextual information with LLM, it can then provide immediate feedback aligned with users’ specific actions, reinforcing skill development, improving task accuracy, and reducing cognitive load.

However, participants also raised concerns about the privacy implications of embedding a camera into the system. While multi-modal systems have been explored in prior research to enhance cooking assistance~\cite{hamada2005cooking, nouri2019supporting, sato2014mimicook}, integrating such features into LLM-based CAs introduces unique challenges. Using a camera has raised privacy concerns, particularly in home settings~\cite{naeini2017privacy, h2022monitoring, chalhoub2021did, park2023nobody}. For instance, users may be hesitant to adopt a system that continuously monitors their actions, even if it enhances the functionality.

Future research could explore strategies to mitigate these privacy concerns while maintaining the benefits of multi-modal capabilities. Possible solutions include implementing strict data processing and storage policies, offering transparent explanations of how visual data is used, and enabling users to toggle the camera on or off. Additionally, investigating user acceptance of multi-modal features in LLM-CAs compared to traditional CAs could provide insights into balancing enhanced functionality with user comfort and trust. Solving these challenges could enable multi-modal LLM-CAs to enhance collaboration while maintaining users' privacy.

\subsection{Problem of Hallucination in LLM-Generated Content}
\label{hallucination}
Hallucination, where LLMs generate nonsensical, irrelevant, or unfaithful content to the user input, remains a significant problem~\cite{huang2023survey,ji2023survey,bang2023multitask}. These inaccuracies are concerning in various applications where incorrect information could lead to serious consequences.
Prior research has explored the use of LLMs and identified the issue of hallucination in diverse scenarios, such as academic research~\cite{kapania2024m}, coding~\cite{tian2024codehalu}, medical~\cite{vishwanath2024faithfulness}, and data annotation~\cite{wang2024human}.  For example,~\citet{kapania2024m} conducted a study on using LLMs for HCI research, where participants expressed concerns that relying on LLMs during the paper writing process could result in misinformation or incorrect references, potentially leading to significant issues in academic work.
Another related challenge is sycophancy, where LLMs generate responses based on users’ beliefs rather than objective facts~\cite{sharma2023towards}.  When using LLM, the ability to engage in multi-round conversations allow the system to accumulates the conversation and uses it as context for future interactions~\cite{huang2023survey}. \citet{manzini2024code} found sycophancy could preventing users from critically assessing their own assumptions that can negatively impact human-AI interaction.

In our cooking experiment, we also observed participants encountering hallucination-related issues. When participants faced suggestions that did not align with their understanding of the recipe from the YouTube video, they would pause and seek clarification, often by rephrasing their query or expressing doubt about the system’s response. Participants’ prior knowledge of how cooking steps should look played a critical role in their ability to identify and question hallucinated responses from \systemname. Since users rely on accurate, step-by-step guidance during cooking, hallucinations in LLM-generated instructions can present significant challenges. For instance, if a user follows a suggested adjustment for ingredient measurements but later tastes the dish and encounters an unexpected taste. Unlike visual interactions, where users can check progress by inspecting a sauce’s consistency, hallucinated suggestions from LLMs can mislead users without visual cues for validation.  This reliance on system guidance increases the potential for user errors, especially when users have no immediate way to verify the correctness of the LLM’s responses. However, participants’ prior knowledge enabled them to detect inconsistencies between their expectations and the system’s guidance. This observation aligns with existing research on tasks like data annotation~\cite{wang2024human}, where human evaluation is critical in identifying and mitigating hallucinated outputs from LLMs.

\subsection{Inclusive Accessibility}
\label{accessibility}

In this study, we conducted experiments with a general population and discovered that LLM-CAs offer multiple advantages over conventional cooking assistance methods. Key benefits include the ability to request information beyond the recipe and more adaptive support aligned with the natural flow of cooking. Beyond general use, we recognize that these capabilities could further support marginalized populations who are often excluded by conventional cooking tools for cooking assistance, offering them greater accessibility and support.

The potential of LLM-CAs to support diverse user group has been demonstrated in prior research across various domains. Conventional tools often fail to support individuals who face challenges with complex interfaces, visual demands, or rigid, linear instructions~\cite{li2020older}. By enabling conversational, hands-free, and context-aware interactions, LLM-CAs make it easier for users to stay engaged in the task while reducing cognitive load. For example,~\citet{kaniwa2024chitchatguide} leveraged LLMs to provide conversational guidance for visually impaired individuals navigating a shopping mall. Participants reported that the system offered easy access to various information, enabled them to ask follow-up questions, and had the ability to plan visit tours, leading to more effective exploration. Similarly, ~\citet{gorniak2024vizability} utilized LLMs to facilitate conversational interaction for visually impaired users to navigate visual data using voice commands. Beyond navigation and visual data interaction, LLM-CAs have also been leveraged in healthcare applications, such as supporting communication between older adults and their providers~\cite{yang2024talk2care}. These LLM-based applications show flexibility and adaptability in supporting diverse accessibility needs across multiple domains.

Given the ample evidence in the literature regarding the potential of LLM-CAs in various domains, our findings suggest that LLM-CAs can also support marginalized groups in cooking contexts. The flexibility and responsiveness observed during our cooking experiments underscore this potential. For instance, older adults frequently rely on technology to track their cooking progress but often struggle to switch between tasks without external support~\cite{kuoppamaki2021designing}. Age-related cognitive or physical changes may further complicate the process of following multi-step instructions, increasing the need for real-time, accessible guidance~\cite{jaroslawska2021age}. LLM-CAs address this challenge through their conversational adaptability, allowing users to request clarification or task-specific guidance at any point during the cooking process. In addition, individuals with visual impairments face additional barriers when using traditional cooking tools, which often rely heavily on text, images, or video-based instructions~\cite{li2024recipe}. Navigating these visually demanding formats can be difficult, especially when precision is required for cooking steps like ingredient measurements or timing. LLM-CAs offer advantage through voice-based interaction, enabling visually impaired users to access recipes, receive detailed verbal instructions, and clarify steps without relying on visual cues. By providing adaptive, real-time guidance, LLM-CAs demonstrate their potential as tools for enhancing independence and inclusivity in cooking tasks, particularly for those with unique accessibility needs.

\subsection{Limitations of the Study}
\label{limitation}
We acknowledge a few limitations in our system and exploratory lab study. While our system evaluation in Section 4.2 did not highlight prominent issues caused by LLM, we acknowledged that certain notorious problems of LLM, such as hallucination, could potentially affect users' experience with \systemname.
Moreover, while we chose cooking as an example due to its representative nature as a sequential but complex real-time task where CAs are already extensively used, we still realized its limitation and tried not to over-generalize our findings to other scenarios, especially considering our small sample size (n=12). Future research could explore the possibility of applying these design implications in different scenarios to test their generalizability and further tailor the results to various use cases in a large-scale study.

Our study focused on a salad-making scenario to minimize safety concerns. As described in the methods section, we selected a relatively complex salad recipe with multiple food preparation and measurement steps to mimic a more comprehensive cooking experience. Simultaneously, we devised a streamlined process for transitioning from recipes to system prompts, aiming to maximize the scalability of our methods. However, the limitation of conducting experiments on only one recipe might still restrict the generalizability of our findings, especially some recipes might involve different cooking processes, for instance, those requiring a stove. Consequently, we aim to articulate the limitations of our experiment clearly. Future research is encouraged to explore various recipes to further enhance the scope of investigation.

In the previous section, we discussed the need for feature discovery of LLM-CA. In our study, we conducted a brief tutorial session beforehand to demonstrate some example questions like ``What is the first step?'', ``What if I don't have chicken, what should I do?'', and ``What did I just ask?''. Despite the training session, users might still require some trial and error to discover their preferred way of using and communicating with the system, which we summarized as one design implication, as well as pointed out as one of the limitations of our experiment.

Lastly, since our primary focus was not on quantitative results, and we did not conduct a comparative study that would provide a baseline for analysis, we mainly used those results to verify our system's basic usability and general user performance. Future work could involve comparative studies to assess the effectiveness of such systems across various dimensions.

\section{Conclusion}
In this study, we explored users' experiences, thoughts, and expectations while interacting with an LLM-based CA system and synthesized design implications for future systems. To achieve this, we conducted a mixed-methods exploratory study with 12 participants and asked them to complete a salad recipe with assistance from our system. We then examined their experiences using surveys, interviews, and interactive logs. Our findings revealed that users quickly adapted to the LLM's capabilities to assist their cooking practices, including asking for extensive information, requesting personalized and context-aware assistance, and dynamically planning their tasks. However, users also expressed the desire for the system to facilitate more natural and oral conversations. Additionally, participants wanted to be more involved in the decision-making process of the CA, suggesting a potential shift in their perception of the system from a tool to a personal assistant and even a partner. Based on these observations, we synthesized design implications for a future LLM-CA.

\begin{acks} 
This work is supported in part by the National Science Foundation (NSF) under award number 	IIS-2141335, IIS-2302730, the National Institutes of Health (NIH) under award number R01AI188576, and the Northeastern University Tier-1 Research Grant. The content is solely the responsibility of the authors and does not necessarily represent the official views of NSF or NIH. 
\end{acks}

\bibliographystyle{ACM-Reference-Format}
\bibliography{bio}

\clearpage
\appendix
\section{Appendix}
\subsection{Prompt Sample for \systemname}
\begin{table*}[h]
\centering
\begin{tabular}{p{.95\linewidth}}
  \toprule
  \textbf{Prompt for \systemname (Part 1: Knowledge Resources) } \\
  \midrule
  RECIPE = \\
    INGREDIENTS FOR CHICKEN AVOCADO MANGO SALAD \\
    - 1 1/2 cups or 1/4 head romaine lettuce, rinsed, chopped and spun dry \\
    - 1/4 lb or 1/2 medium cooked chicken breasts  \\
    - 1/4 mango, pitted, peeled and diced \\
    - 1/4 avocado, pitted, peeled and diced \\
    - 1/8 english cucumber sliced  \\
    - 1/8 thinly sliced small purple onion \\
    - 1/8 cup halved cherry tomatoes \\
    - 1/16 cup chopped cilantro chopped \\
    \\
    STEPS  \\
    - Step 1: Chop the romaine into bite-sized pieces and discard the core.  
    After rinse and spin dry, place it in a large salad bowl.   \\
    - Step 2: Slide chicken into bite size strips and place it over the romaine lettuce.  \\
    - Step 3: Place diced mango in to salad bowl.  \\
    - Step 4: Peel and dice the advocado, then place it on top of the salad bowl.  \\
    - Step 5: Place slices cucumber in to salad bowl.  \\
    - Step 6: Added thinly sliced small purple onion.  \\
    - Step 7: Cut the cherry tomatoes into half and place it on the salad.  \\
    - Step 8: Add chopped fresh cilantro.  \\
    \\
    INGREDIENTS FOR HONEY VINAIGRETTE DRESSING \\
    - 1/8 cup extra virgin olive oil \\
    - 3/4 Tbsp apple cider vinegar \\
    - 1/2 tsp dijon mustard \\
    - 1/2 tsp honey \\
    - 1/4 garlic clove or 1/4 tsp minced garlic \\
    - 1/4 tsp sea salt \\
    - 1/16 tsp black pepper, or to taste \\
    \\
    - Step 9: Combine the Honey Vinaigrette Dressing Ingredients in a mason jar, 
    first add olive oil. \\
    - Step 10: Add apple cider vinegar, Dijon mustard and honey \\
    - Step 11: Add garlic, sea salt and black pepper \\
    - Step 12: Cover tightly with lid and shake together until well combined.  \\
    - Step 13: Drizzle the salad dressing over the chicken mango avocado salad, adding it to taste. 
    \\
  \bottomrule
\end{tabular}

\caption{ Prompt for \systemname of Knowledge Resources. }
\label{tab:prompt1}
\end{table*}
\begin{table*}[!h]
\centering

\begin{tabular}{p{.95\linewidth}}
  \toprule
  \textbf{Prompt for \systemname (cont. Part 2: Instructions) } \\
  \midrule
  INSTRUCTIONS =  \\
    Your main task is to help guiding user to make the chicken avocado mango salad step 
    by step based on the recipe provided delimited by triple backticks.  \\
    The recipe is for 1 person. \\
    There are 2 parts of this recipe: the salad part and the dressing part. \\
    Please follow these steps to guide user by answering the customer queries.  \\
     \\
    1:   First decide whether the user is asking a question about a specific 
    ingredients or recipe steps or other. When user ask for next step, assume user is about to perform that step.  \\
    Once the dressing steps are finished or all the ingredients are placed, the entire recipe is complete, and no more futher steps since all salad and dressing steps and ingredients covered. Congratulate 
    user and tell user all the steps are complete. \\
     \\
    2:  If the user is asking about overall ingredients, for example: how to make 
    the dressing. Respond with all the ingredients without measurements, for 
    example: The ingredients for chicken avocado mango salad are romaine 
    lettuce, chicken breasts. Do not respond: The ingredients for chicken avocado 
    mango salad are 1 lb or 2 medium cooked chicken breasts and 6 cups or 1 head 
    romaine lettuce. \\
     \\
    3:  If the user is asking about one specific ingredients. Identify whether 
    the ingredients is for the salad or the salad dressing, then respond corresponding 
    ingredients with measurement. For example: 1/2 thinly sliced small purple 
    onion is needed for the salad. \\
     \\
    4:  If the user is asking about specific steps, 
    identify what step of the recipe the user is working on, then respond with 
    short, clear and easy to follow instructions. \\
     \\
    5:  Respond to user with summarizing the response from steps above in 30 words or less.
    Please response in complete sentence. Please aim to be as helpful, creative, 
    friendly, and educative as possible in all of your responses.  \\
    Do not use any external recipe in your responses. \\
    For question not related to this recipe, try your best to answer it. \\
  \bottomrule
\end{tabular}
\caption{ Prompt for \systemname of Instructions }
\label{tab:prompt2}
\end{table*}

\clearpage
\subsection{Questionnaire Results}
\label{appendix:questionnaire}
\begin{table}[h!]
  \footnotesize
  \centering
  \begin{tabular*}{0.99\linewidth}{@{\extracolsep{\fill}}l l}
    \makecell[cl]{\textbf{Voice Usability Scale(VUS) Questions}}
    & \makecell[cl]{\textbf{Mean(SD)}}\\
    \midrule
    \makecell[cl]{\textbf{Usability}}\\
        \makecell[cl]{I thought the information provided by the \systemname was not relevant to what I asked.} 
        & \makecell[tl]{2.333(1.371)}\\
        \makecell[cl]{I thought the \systemname had difficulty in understanding what I asked it to do.}    
        & \makecell[tl]{2.083(0.515)}\\
        \makecell[tl]{I found the \systemname difficult to use.}  
        & \makecell[tl]{1.667(0.985)}\\
    \makecell[cl]{\textbf{Affective}}\\
        \makecell[cl]{I felt the \systemname enabled me to successfully complete my tasks when I required help.}    
        & \makecell[tl]{4.500(0.522)}\\
        \makecell[cl]{The \systemname had all the functions and capabilities that I expected it to have.}  
        & \makecell[tl]{4.333(0.985)}\\
        \makecell[tl]{I felt the response from the \systemname was sufficient.}  
        & \makecell[tl]{3.833(1.115)}\\
        \makecell[cl]{Overall, I am satisfied with using the \systemname.}    
        & \makecell[cl]{4.333(0.888)}\\
    \makecell[cl]{\textbf{Recognizability \& Visibility}}\\
        \makecell[cl]{I thought the response from the \systemname was easy to understand.}    
        & \makecell[tl]{4.333(0.888)}\\
        \makecell[cl]{I found it difficult to customize the \systemname according to my needs and preferences.}  
        & \makecell[tl]{2.167(1.030)}\\

    \bottomrule
  \end{tabular*}
\caption{The questions of Voice Usability Scale(VUS) and results in the format of Mean (Standard Deviation)}
\label{tab:vus_result}
\vspace{-0.35in}
\end{table}

\begin{table}[h!]
  \footnotesize
  \centering
  \begin{tabular*}{0.99\linewidth}{@{\extracolsep{\fill}}l l}
    \makecell[cl]{\textbf{Explainable AI (XAI) Questions}}
    & \makecell[cl]{\textbf{Mean(SD)}}\\
    \midrule
    \makecell[tl]{I am confident in the \systemname. I feel that it works well.}    
    & \makecell[tl]{4.250(0.866)}\\
    \makecell[tl]{The outputs of the \systemname are very predictable.}    
    & \makecell[tl]{4.250(0.965)}\\
    \makecell[cl]{I feel safe that when I rely on \systemname that I will get the right response.}    
    & \makecell[cl]{3.917(0.996)}\\
    \makecell[tl]{\systemname is efficient in that it works very quickly.}    
    & \makecell[tl]{4.000(1.348)}\\
    \makecell[tl]{\systemname can better help me than the recipes in other formats.}    
    & \makecell[tl]{3.917(1.443)}\\
    \makecell[tl]{I like using \systemname for cooking instructions.}    
    & \makecell[tl]{4.333(1.073)}\\
    \bottomrule
  \end{tabular*}
\caption{The questions of Explainable AI (XAI) survey and results in the format of Mean (Standard Deviation)}
\label{tab:xai}
\vspace{-0.35in}
\end{table}

\begin{table}[h]
  \footnotesize
  \centering
  \begin{tabular*}{0.99\linewidth}{@{\extracolsep{\fill}}l l}
    \makecell[cl]{\textbf{NASA-TLX Questions}}
    & \makecell[cl]{\textbf{Mean(SD)}}\\
    \midrule
    \makecell[tl]{How mentally demanding was it to interact with \systemname?}    
    & \makecell[tl]{2.583(1.240)}\\
    \makecell[tl]{How physically demanding was it to interact with \systemname?}    
    & \makecell[tl]{1.917(1.379)}\\
    \makecell[tl]{How hurried or rushed was it to interact with \systemname?}    
    & \makecell[tl]{2.000(0.853)}\\
    \makecell[tl]{How successful were you in communicating with \systemname?}    
    & \makecell[tl]{3.750(0.965)}\\
    \makecell[tl]{How hard did you have to try to communicate with \systemname?}    
    & \makecell[tl]{2.667(1.073)}\\
    \makecell[cl]{How insecure, discouraged, irritated, stressed, and annoyed were you communicating\\ with \systemname?}    
    & \makecell[cl]{2.167(1.267)}\\
    \bottomrule
  \end{tabular*}
\caption{The questions of NASA-TLX and results in the format of Mean (Standard Deviation)}
\label{tab:nasatlx}
\vspace{-0.35in}
\end{table}

\begin{table}[h!]
  \footnotesize
  \centering
  \begin{tabular*}{0.99\linewidth}{@{\extracolsep{\fill}}l l l}
    \makecell[cl]{\textbf{Exploration Questions}}
    & \makecell[cl]{\textbf{Mean(SD)} \\ \textbf{Pre Study}}
    & \makecell[cl]{\textbf{Mean(SD)} \\  \textbf{Post Study}}\\
    \midrule
    \makecell[cl]{I thought the current voice assistants could engage in fluent and human-like \\conversations.}
    & \makecell[cl]{3.500(0.798)}
    & \makecell[cl]{4.167(0.835)}\\
    \makecell[cl]{I thought the current voice assistant has the ability to remember and refer back\\ to previous parts of a conversation.}    
    & \makecell[cl]{3.167(0.937)}
    & \makecell[cl]{4.167(0.718)}\\
    \makecell[cl]{I thought the current voice assistant allowed asking follow-up questions that\\ relate to the ongoing conversation.}   
    & \makecell[cl]{3.417(0.996)}
    & \makecell[cl]{4.333(0.778)}\\
    \makecell[cl]{I thought the current voice assistant can seamlessly integrate into my daily activities.}    
    & \makecell[cl]{3.333(0.985)}
    & \makecell[cl]{3.917(0.900)}\\
    \makecell[cl]{I thought the current voice assistant can actively collaborate with me on different tasks.}    
    & \makecell[cl]{3.167(0.835)}
    & \makecell[cl]{3.833(1.193)}\\
    \bottomrule
  \end{tabular*}
\caption{The five supplementary questions we created to understand users’ perspectives. Pre-study and post-study results in the format of Mean (Standard Deviation)}
\label{tab:explore}
\end{table}

\received{July 2024}
\received[revised]{December 2024}
\received[accepted]{March 2025}

\end{document}